\documentclass[aps,pre,reprint,superscriptaddress,amsmath,showpacs]{revtex4-1}
\usepackage[breaklinks,colorlinks=true]{hyperref}
\usepackage{graphicx}
\usepackage{amssymb}
\usepackage{amsmath}
\usepackage{verbatim}
\usepackage{chngcntr}

\begin{document}
%%Title
\title{Jamming and condensation in one-dimensional driven flow}

\author{Hyungjoon Soh}
\email[Present address: Kakao, Pangyoyeok-ro 235, Seongnam, Republic of Korea 13494; ]{jay.mini@kakaobrain.com}
\affiliation{Department of Physics,
	Korea Advanced Institute of Science and Technology, Daejeon
	34141, Republic of Korea}	

\author{Meesoon Ha}
\email[Corresponding author; ]{msha@chosun.ac.kr}
%\email[]{msha@chosun.ac.kr}
\affiliation{Department of Physics Education, Chosun University,
	Gwangju 61452, Korea}

\author{Hawoong Jeong}
\email[ ]{hjeong@kaist.edu}
\affiliation{Department of Physics,
	Korea Advanced Institute of Science and Technology, Daejeon
	34141, Korea}
\affiliation{Institute for the
	BioCentury, Korea Advanced Institute of Science and Technology,
	Daejeon 34141, Korea}

\date{\today}

\begin{abstract}
We revisit the slow-bond (SB) problem of the one-dimensional (1D) totally asymmetric simple exclusion process (TASEP) with modified hopping rates. In the original SB problem, it turns out that a local defect is always relevant to the system as jamming, so that phase separation occurs in the 1D TASEP. However, crossover scaling behaviors are also observed as finite-size effects. In order to check if the SB can be irrelevant to the system with particle interaction, we employ the condensation concept in the zero-range process. The hopping rate in the modified TASEP depends on the interaction parameter and the distance up to the nearest particle in the moving direction, besides the SB factor. In particular, we focus on the interplay of jamming and condensation in the current-density relation of 1D driven flow. Based on mean-field calculations, we present the fundamental diagram and the phase diagram of the modified SB problem, which are numerically checked. Finally, we discuss how the condensation of holes suppresses the jamming of particles and vice versa, where the partially-condensed phase is the most interesting, compared to that in the original SB problem. 
\end{abstract}

\pacs{02.50.-r, 05.40.-a, 64.60.-i, 89.75.Da}

%----------------------------------------------------------------------------
%02.50.-r  Probability theory, stochastic processes, and statistics
%89.75.Hc Complex systems: networks and genealogical trees
%05.40.-a Fluctuation phenomena, random processes, noise, and Brownian motion 
%05.60.Cd Classical transport
%64.60.-i   General studies of phase transitions
%89.75.Da	Systems obeying scaling laws
%----------------------------------------------------------------------------

\maketitle

\section{Introduction}
\label{sec:intro}

%Introduction of driven diffusive system, KPZ solution, TASEP
Driven diffusive systems are ubiquitous in real-world phenomena with various scales, from active transport in cell proteins~\cite{Chowdhury2005, Bressloff2013, Neri2013b} to large traffic networks~\cite{Helbing2001, Chowdhury2000, Embley2009, Schadschneider2000}. As the simplest modeling of such system, the stochastic (noisy) Burger's equation~\cite{Forster1976} is often employed, which is also known as the Kardar-Parisi-Zhang (KPZ) equation~\cite{Kardar1986}. Most recently, the detailed statistical properties of the one-dimensional (1D) KPZ equation has been exactly solved by mathematicians, in terms of the random matrix formalism~\cite{Corwin2011, Quastel2011}. The simplest one of the models that belong to the 1D KPZ university class, is the \textit{totally asymmetric simple exclusion process} (TASEP)~\cite{Derrida1992}. It is well-established that the TASEP is a prototype model of nonequilibrium driven flow, and its stationary solutions with various boundary conditions are presented by a matrix-product ansatz~\cite{Derrida1992,Blythe2007}.

%Asymmetric fundamental diagram
In the ordinary TASEP, the current-density relation is symmetric with a single maximum in the 1D TASEP, which is due to the particle-hole symmetry. When the hopping rate is modified with particle interaction, the symmetry is broken in the fundamental diagram of flow (the current-density relation). Similarly, a local defect indeed also changes the shape of the fundamental diagram. Such an example is the slow-bond (SB) problem~\cite{MHa2003,Costin2012,Schmidt2015,HSoh2017}. In the SB problem, the driven flow in the middle of the system becomes slow as the hopping rate at the SB is reduced, compared to that at normal bonds. The most interesting question of the SB problem is ``whether the SB effect is always relevant to the system so that the fundamental diagram is changed.'' This has also been speculated in various studies, such as slow combustion of paper with a local columnar defect~\cite{Myllys2003}, the modified KPZ growth models in random media~\cite{Kandel1992,HSSong2006}, directed polymer in random media~\cite{JHLee2009}, last passage percolation~\cite{Basu2014}, and junctional defect of networks with TASEP links~\cite{Neri2011,YBaek2014}. 

The possibility of the SB-irrelevant phase was proposed in the ordinary TASEP by numerical simulations~\cite{MHa2003} and experiments~\cite{Myllys2003}, but it was hardly proven since nontrivial crossover scaling behaviors exist as well as boundary effects. As the SB strength gets close to 1, the localization of the queue occurs in finite systems. However, it turns out that such a phenomenon is attributed to the finite-size effect~\cite{HSoh2017}, consistent with analytic arguments proposed by Costin and coworkers~\cite{Costin2012}. 

In this paper, we employ the hopping rate of the zero-range process (ZRP)~\cite{Spitzer1970,Godreche2003,Evans2003} as particle interaction in the TASEP with a SB at the middle of the system~\footnote{The ZRP is an exact mapping of the periodic TASEP without the slow bond.}. In the context of the ZRP dealing with mass transport, the hopping rate depends only on the mass at the chosen site. 
The most interesting phenomenon in the ZRP is the condensation of mass at a single site, which occurs when the interaction parameter gets positively larger than the certain value under the circumstances. In the TASEP language, the condensation of holes is the particle-hole segregation. The SB induces the queue of particles so that the bulk density is not single-valued anymore even far from the SB.

In particular, we investigate the interplay of the SB effect and particle interaction in the current-density relation, in terms of the modified TASEP with periodic boundary conditions. Considering modified ZRP-type hopping rates at all bonds, in the modified SB problem, we pose the question, ``\textit{Is it possible that the condensation can suppress the queue by the SB effect and/or vice versa?}''. 
To answer this question, we focus on the fundamental diagram of current-density relations as well as the phase diagram. Based on the mean-field (MF) calculations of the current-density relation, we suggest a possible ``bulk'' density and propose the phase diagram in the modified SB problem, which is compared to numerics. In the regime where the correlation length does not globally diverge, our numerical results also show that the system separates into two homogeneous subsystems with the same current but different bulk densities. 

However, due to the particle conservation of the periodic TASEP, the system may not be well separated when all of the allowed densities are even lower (higher) than the total density of the system. Applying MF approximations, we seek the marginal phase boundaries as the function of the SB factor and the interaction parameter, which are numerically checked. Moreover, we discuss the partially condensed phase in the strong SB regime with possible physical arguments. 

The rest of the paper is organized as follows: In Sec.~\ref{sec:model}, we describe the modified SB problem, in terms of the TASEP with ZRP-type modified hopping rates as well as the SB, where physically relevant quantities are denoted as two control parameters vary. In Sec.~\ref{sec:PD}, we present the MF approximations of the phase diagram and its marginal phase boundaries, in the context of the fundamental diagram of the modified SB problem, where we present four different phases. Extensive Monte Carlo (MC) numerical simulation results are provided for the comparison with MF results in Sec.~\ref{sec:numerics}, where finite-size effects are also carefully tested. Finally, in Sec.~\ref{sec:summary}, we conclude this paper with the summary of our findings and some remarks. Fo more additional information with additional figues, we provide Appendices~\ref{appendix1}--\ref{appendix3}.

\section{Model}
\label{sec:model}

\begin{figure}[]
	%Figure 1 : Schematic diagram
	\centering
	\includegraphics[width=\columnwidth]{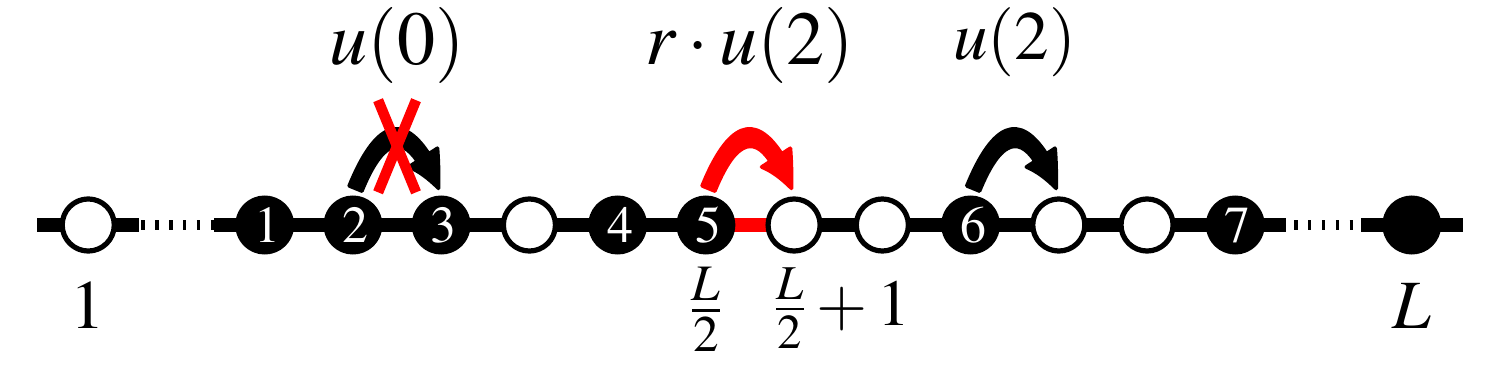}
	\caption{\label{fig:model} The modified TASEP is schematically illustrated, where the numbers above the arrows indicate hopping rates, and the site indices are shown at the bottom. The hopping is forbidden due to hard-core repulsion, which is shown as a red cross, and the hopping at the SB is highlighted as the different color (red) arrow.}
	%\caption{\label{fig:model} The modified TASEP is schematically illustrated with its mapping to the ZRP. The numbers above the arrows indicate hopping rates, and the site indices are shown at the bottom. The hopping is forbidden due to hard-core repulsion, which is shown as a red cross and the hopping at the SB is highlighted as the different color (red) arrow. Note that for the case of $r = 1$ with periodic boundary conditions, the mapping is exact.}
\end{figure}

We consider the modified TASEP in a 1D lattice of even $L$ sites as shown in Fig.~\ref{fig:model}, where each site is either occupied by at most one particle or vacant only at time $t$, $n_x(t) = \{0, 1\} ~(1\le x\le L)$. The hopping rate depends on the distance from the chosen particle up to the nearest particle in the hopping direction. Employing periodic boundary conditions, $n_{_{L+1}}= n_{_1}$ and the total number of particles, $N = \rho_{_0} L$, where $\rho_{_0}$ is fixed as the total density of the system. Finally, we place the SB to be between $x=L/2$ and $x=L/2+1$, where the hopping probability is suppressed as a factor $r\in[0,1)$. Without loss of generality, the case of $\rho_{_0}$=1/2 is chosen and compared with the original one~\cite{HSoh2017} as the modified SB problem. 

By definition, the average occupancy and the average interparticle distance has the following relation if the system is homogeneous: 
\begin{align}
[n_x]\equiv\sum_{x=1}^L \frac{n_x}{L},\ [\ell_y]_{_N}\equiv\sum_{y=1}^N \frac{\ell_y}{N};~~[n_x]=\frac{1}{1 + [\ell_y]_{_N}}.
\end{align}

At each time step, particle configurations in the modified TASEP are updated [see Fig.~\ref{fig:model} (TASEP-type) and its possible mapping in Fig.~\ref{fig:mapping-zrp} (ZRP-type) in Appendix~\ref{appendix1}]: 
\begin{enumerate}
	\item{Choose one among $N$ particles at random, \textit{e.g.},  the $i$-th particle at site $x_i$.}
	\item{The $i$-th particle hops to the next site, $x_i+1$, with probability $q(\ell_i)$, where $\ell_{i} = x_{i+1}-x_{i}-1$ (the distance up to the site of the nearest particle in the target direction):
		\begin{align}
		q(\ell_i) = 
		\begin{cases}
		0 & (\ell_i = 0), \\
		u(\ell_i)/{u_{_{\rm max}}} & (\ell_i \geq 1),
		\end{cases}
		\label{eq:q}
		\end{align}
		where  $u(\ell_i) = \left( 1+\frac{b}{\ell_i} \right)$,  $b$ is an interaction parameter, and  $u_{_{\rm max}}$ is the maximum hopping rate.}
	\item{If $x_i=L/2$, the particle has to get through the SB, so that the hopping is suppressed by a factor $r(<1)$, namely the SB factor.}
\end{enumerate}

For the modified TASEP with the hopping rates of Eq.~(\ref{eq:q}), $u_{_{\rm max}}$ rescales hopping rates into probabilities by the maximum rate to 1, which depends on the sign of $b$: 
\begin{align}
u_{_{\rm max}} = 
\begin{cases}
1+b & \mbox{for $b>0$ (attractive)},\\
1+b/L(1-\rho_{_0}) &\mbox{for $b<0$ (repulsive)}. 
\end{cases}
\end{align}
Note that the case of $b=0$ corresponds the ordinary TASEP, and $q(\ell_i=0)=0$ implies hard-core repulsion (exclusion). When $b < 0$, particles prefer to be equally spaced, which drives the system to have almost the same value of $\ell_i$, irrespective of $i$. In the periodic system with the fixed density, this force acts particles effectively to repel each other. On the other hand, when $b > 0$, particles prefer to be close each other. When this attractive force is greater than the critical strength, $b > b_c$, the system segregates particles and holes (vacancies) to form macroscopic condensate of holes.

%%% Add arguments about modified ZRP%%%The distance-dependent hopping rate provides the standard mapping of the modified TASEP onto the ZRP without SB, the dynamics becomes complex and unable to express in terms of ZRP for $r<1$ (see Appendix~\ref{appendix2} for more details). Still, the inter-particle distance distribution $P(\ell)$ is a key observable of the condensation of holes. 

The most relevant physical quantities of driven flow are the bulk density $\rho$ and the current of the system $J$ because the current-density relation is the fundamental diagram of driven flow $J(\rho;b)$ that determines the detailed phase structure of the model-dependent phase diagram. The local current at the bond $(x,x+1)$ due to the movement of the $y$-th particle located at site $x$, $J_{x,x+1}$, is denoted as
\begin{align}
J_{x,x+1} = \langle n_x u(\ell_y) \rangle,
\label{eq:J-rho}
\end{align}
where $u(\ell_y)=(1+b/\ell_y)$ is the hopping rate of the $y$-th particle with $\ell_y$. 
In the stationary state of the homogeneous system ($J_{x,x+1}\approx J, \langle n_x\rangle\approx\rho, \ell_y\approx \ell$), the current $J$ of the modified TASEP can decouple with the bulk density and the average hopping rate as $J_{_{\rm MF}} = \rho \langle u(\ell) \rangle$, where the mean-field (MF) approximations are valid as described in Appendix~\ref{appendix2}. It is worthwhile to mention that $J$ is distinguished from the conventional TASEP current $\widetilde{J}_{_{\rm MF}} = J_{_{\rm MF}} / u_{_{\rm max}}=\rho\langle q(\ell)\rangle$.

However, the SB leads to jamming, so that the system becomes inhomogeneous. The case of $b=0$ is the well-known SB problem~\cite{Janowsky1992,MHa2003,Costin2012,Basu2014,Schmidt2015,HSoh2017}, where the main issue was the possibility of the homogeneity if the SB effect is weak enough to be irrelevant in the fundamental diagram. Although it looks possible in finite systems due to crossover scaling caused by finite-size effects, the SB effect is always relevant~\cite{HSoh2017}.

In this paper, we pose the following question: \textit{Can particle interactions suppress the SB effect, so that the queue by the SB can be localized in the thermodynamic limit, unlike the original SB problem?}

\begin{figure*}[]
	%Figure 2 : Phase diagrams
	\centering
	\includegraphics[width=\textwidth]{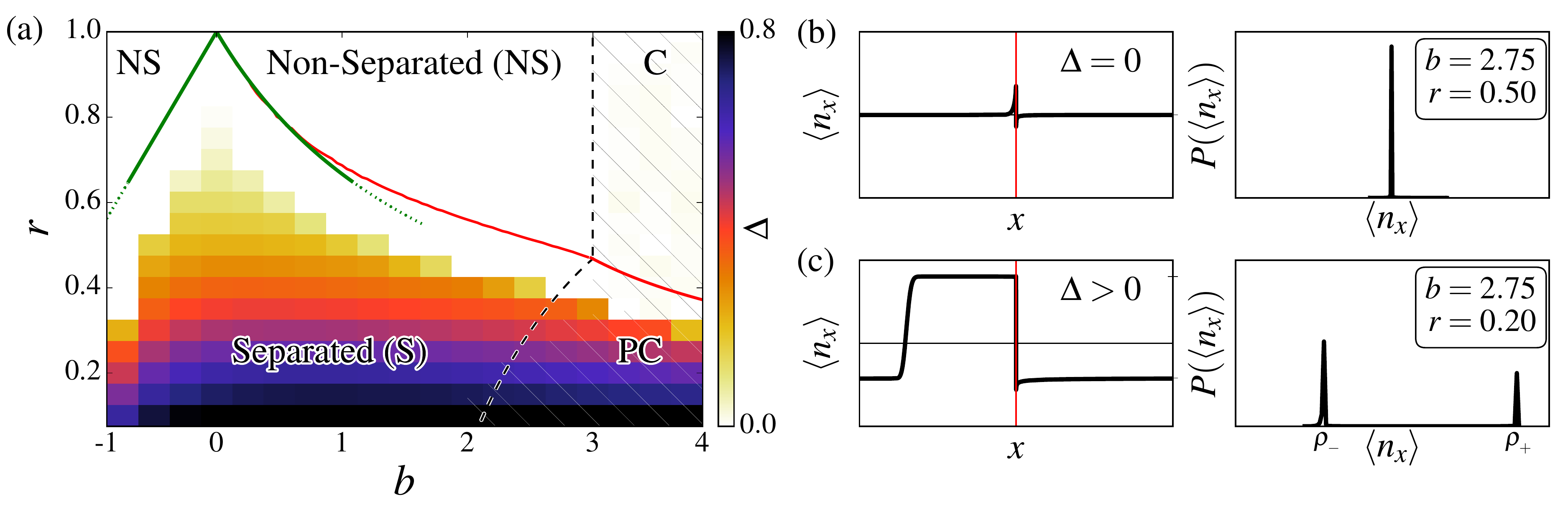}
	\caption{\label{fig:PhaseDiagram} (a) In the $b-r$ plane, the phase diagram of the modified SB problem is drawn as a heatmap, in terms of the density difference $\Delta = \rho_{_+}-\rho_{_-}$. MF phase boundaries are drawn for the NS-S from Eq.~\eqref{eq:PB-NS-S} (red, solid line) up to $0<b<3$, and the C-PC from Eq.~\eqref{eq:PB-C-PC} (red, solid line) for $b>3$, and the S-PC from Eq.~\eqref{eq:PB-S-PC} (dashed line). The NS-C boundary is obtained from Eq.~\eqref{eq:PB-NS-C} (dashed line). Around $b = 0$, the NS-S-NS boundaries are drawn from Eq.~\eqref{eq:PB-b0} (green, solid and dotted lines). The shaded region presents with $J/\rho_{_-} > 0.999$, which denotes condensation of holes in numerical measurement. For examples, we show the NS phase at (b) $(b=2.75, r=0.5)$ with $\Delta = 0$ and the PC phase at (c) $(2.75, 0.2)$ with $\Delta>0$. Numerical data are obtained for $L=2^{16}$ at $T\gg 2L^{3/2}$, averaging over $10^8$ samples with $10^4$ different configurations and $10^4$ different times.}
\end{figure*}
\begin{figure*}[]
	%Figure 3 : Spatiotemporal density profiles
	\centering
	\includegraphics[width=0.95\textwidth]{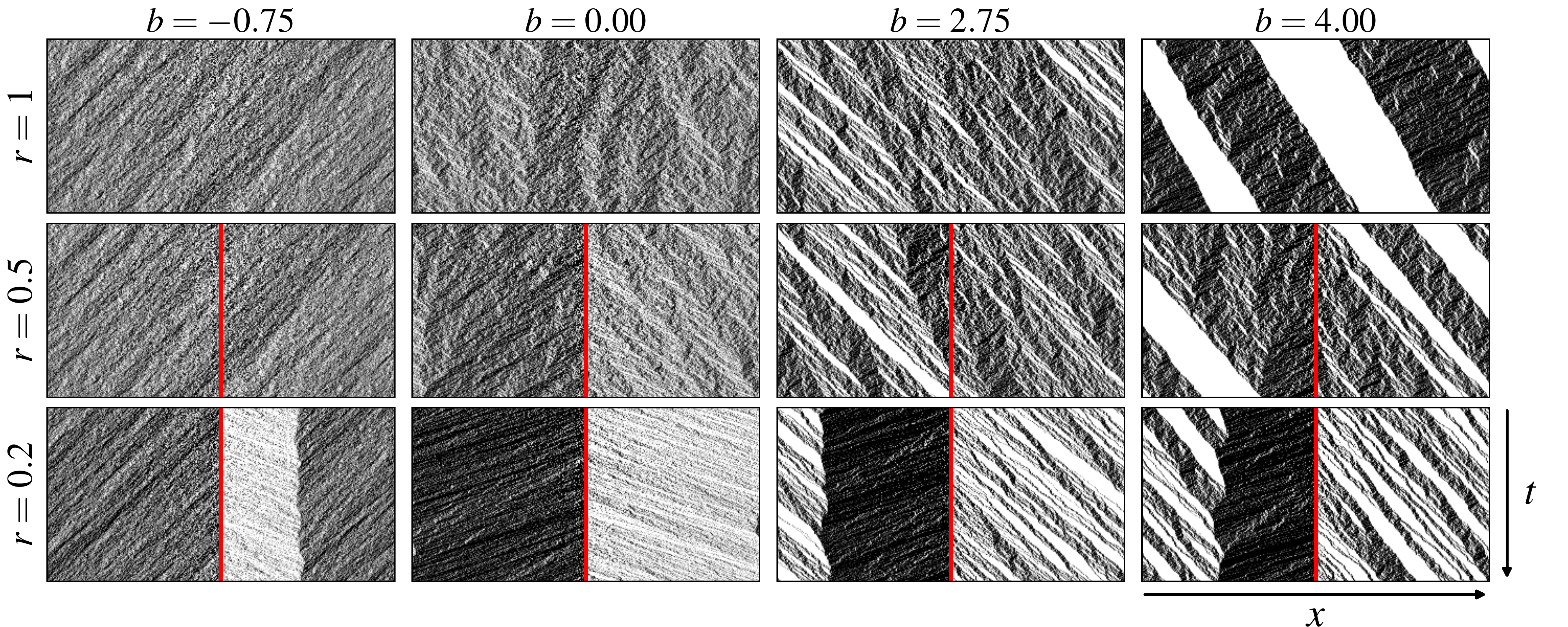}
	\caption{\label{fig:ConfigProfile} Snapshots of spatiotemporal patterns of $L=2^{10}$ (horizontal length of each box) and $T=768$ (vertical length of each box) are plotted at every consecutive $\Delta t=4~\mbox{MC}$ time step in the steady states $(t \gg 2L^{3/2})$ for various cases. Dots represent particles, and the SB is highlighted for the case of $r<1$ as the vertical line in the middle of the pattern with a different color (red). In each pattern, time elapses from top to bottom, and the direction of particle hopping is to the right. Each column represents the different type of particle interaction: Repulsive ($b=-0.75$), neutral ($b = 0$), attractive ($b=2.75$), and strongly attractive ($b = 4.00$), respectively. Each row is classified by the different SB factor; $r = 1.0, 0.5, \mbox{and}~0.2$, respectively. For the PC phase, a typical configuration is presented as the two rightmost in the bottom panel.}
\end{figure*}

\section{Phase Diagram} 
\label{sec:PD}

We present a phase diagram in the modified TASEP with a SB, where we categorize four phases, namely separated (S), non-separated (NS), condensed (C), and partially-condensed (PC). The definition of each phase can be identified by density profiles and inter-particle distance distribution functions. 
While both the NS and C phases are described by a bulk density, the S and PC phases are categorized by two bulk densities. Despite this simple concept, density separation is identified with some caution. Unlike previous studies~\cite{MHa2003,HSoh2017} that used density profile $\langle n_x\rangle$ directly, we cannot use it since the broken particle-hole symmetry is not guaranteed for the functional shape of density profile and the location of bulk boundaries.

A suitable indicator of density separation, the density difference, denotes $\Delta = \rho_{_+}-\rho_{_-}$. Thus, the S/PC phase ($\Delta > 0$) can be distinguished from the NS/C phase ($\Delta = 0$), without loss of generality. 
Figure~\ref{fig:PhaseDiagram} represents (a) the detailed phase diagram of the modified SB problem in the space of the interaction parameter $b$ and the SB factor $r$, and density profiles and the local-density distributions are also presented in (b) $ (b=2.75, r=0.5)$ in the NS phase and (c) $(2.75, 0.2)$ in the PC phase.
For the C phase, the SB is irrelevant as shown at the rightmost plot of the middle panel in Fig.~\ref{fig:ConfigProfile} ($b=4.00, r=0.50$), while, for the PC phase, holes in the low-density region form multiple macroscopic condensates as shown at two rightmost plots of the bottom panel in Fig.~\ref{fig:ConfigProfile} ($b=4.00, r=0.20$). 

In order to measure $\Delta$, we suggest the local density distribution function $P(\langle n_x\rangle)$ for density profiles in the thermodynamic limit:  The contribution by bulk boundaries vanish and local densities fluctuate around the high-density (HD) value $\rho_{_+}(>1/2)$ and the low-density (LD) one $\rho_{_-}(<1/2)$ [see Figs.~\ref{fig:PhaseDiagram} (b) and ~\ref{fig:PhaseDiagram}(c)]. As a result,
\begin{align}
P(\langle n_x\rangle) = c_{_+} \delta_{\rho_{_+}}(\langle n_x\rangle)+ c_{_-} \delta_{\rho_{_-}}(\langle n_x\rangle),
\end{align}
where $\delta_{\rho}(x)$ is a delta function centered at $x=\rho$. Furthermore, vacancies (holes) can form one or more macroscopic condensates. If the size of the condensate scales with the system size as $L^\alpha$, the value of the exponent $\alpha \in (0,1]$ determines C and PC phases. 

In the TASEP language, condensation of holes occurs when the average hopping rate becomes $1$, which is restricted by the front particle in the queue of particles. We denote the NS phase with condensation and $\alpha=1$ as the C phase, and the S phase with partial condensate and $0<\alpha<1$ as the PC phase. 
%"Two-bulk" picture and critical slow-bond strength.
For both S and PC phases with $\Delta>0$, the SB still allows the system to have only two bulk densities at most, because the current-density relation still is a single-peaked function at arbitrary $b$. 

Accordingly, in the thermodynamic limit, the system with density-separated phases has a finite correlation length, and it can be simplified as two homogeneous subsystems in contact. This is quite different from the maximal-current phase in the open TASEP, where the divergent correlation length disturbs the system to have a homogeneous bulk density. As long as the correlation length is finite, density-separated phases are composed of homogeneous subsystems with the equal current. On the other hand, the total density conservation restricts the system not to be split into subsystems with both greater or lesser than the total density, \textit{e.g.}, $\rho_{_0}=1/2$ in this paper. 

When the total density $\rho_{_0}$ is given, the marginal high orlow-density $\rho^*_{_\pm}$ should suffice the following relation: 
\begin{align}
\label{eq:MFCriteria}
J(\rho_{_0}) = J(\rho^*_{_\pm}).
\end{align}
Note that there is the ``forbidden'' density region caused by the conservation of both the total density and the current, where some HD or LD counterpart is not allowed.

Once the SB is considered in the system, the critical SB factor $r^*(b)$ can determine the boundary between S and NS phases as a function of the interaction parameter $b$. At the S-NS phase boundary, we can discuss the queuing transition in the modified SB problem with particle interaction, similar to the original SB problem. The transition between S and NS phases is not simply characterized. As described in the two-bulk picture, particle correlations near the SB competing with density separation leads to the essential singularity-like density jump at $r^*(b)$, which is numerically verified in ordinary TASEP~\cite{HSoh2017} [$r^*(0)=1$ at $b=0$]. 

Approaching the maximal current by $r \rightarrow r^*(b)$, the correlation length grows, and the system deviates from the two-bulk picture, smoothly transiting into the NS phase (see Fig.~\ref{fig:PhaseDiagram}). However, it is not clear if the system has an essential singularity because the density profile decays algebraically at only one side of the SB. 

In Fig.~\ref{fig:ConfigProfile}, we show spatiotemporal patterns as snapshots, where 12 different settings of $(b,r)$ are chosen for $b\in \{-0.75, 0.00, 2,75, 4.00\}$ (from left to right) and $r\in \{1, 0.5, 0.2\}$ (from top to bottom). The ordinary TASEP corresponds the case of $(b=0, r=1)$, where the condensation of holes occurs as $b\to 3$ at $\rho_{_0}=1/2$; thus, the pattern of $(b=4.00, r=1)$ represents the C phase. However, the C phase is shrunk by the PC phase that appears as $r$ gets smaller. Condensation suppressed by jamming of particles behind the SB as long as $b$ is large enough to make a partial condensate of holes.

\begin{figure}[]
	%Figure 4 : Density profiles (Density profile)
	\centering
	\includegraphics[width=\columnwidth]{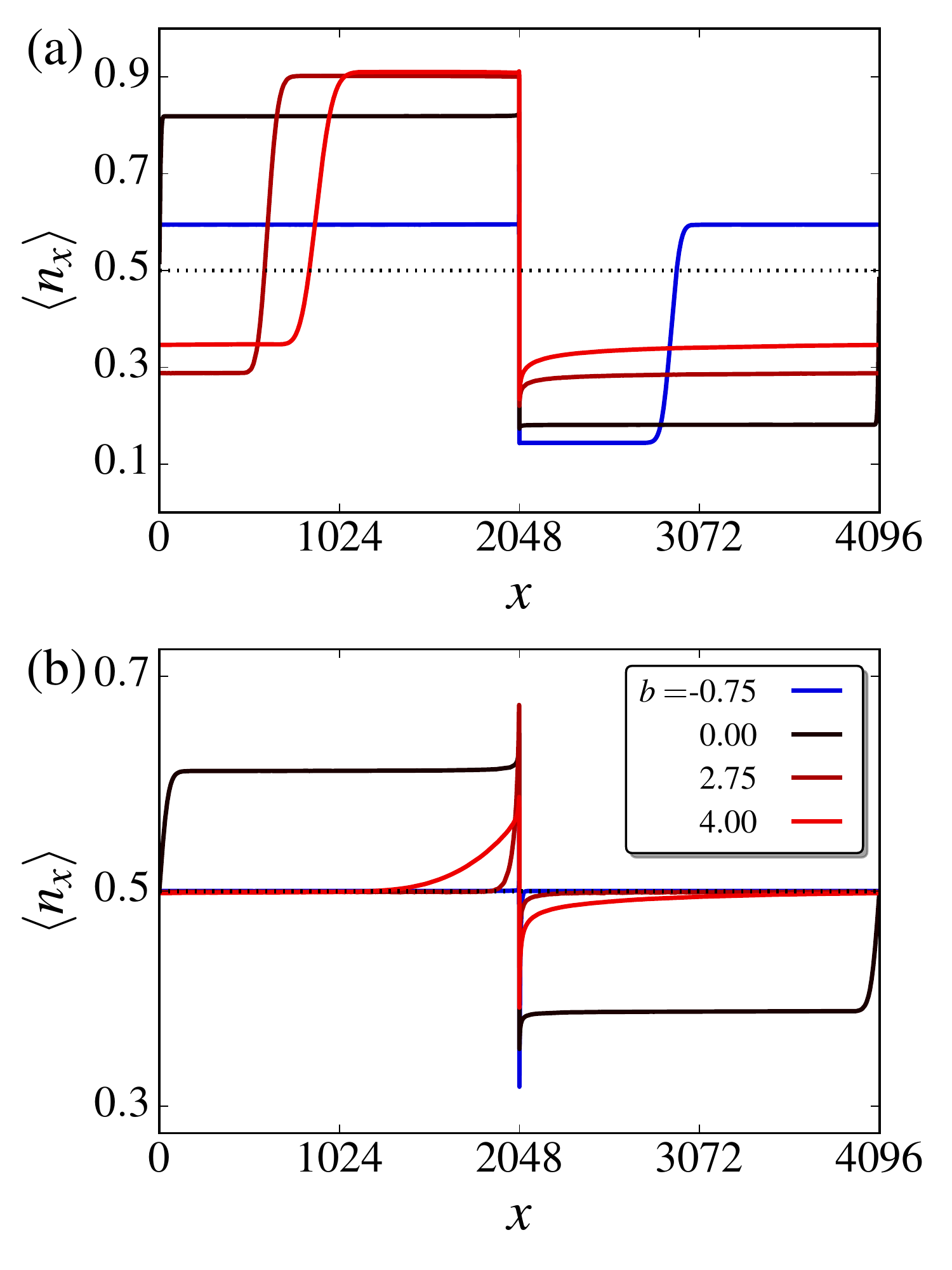}
	\caption{\label{fig:Density} Density profiles are shown as $b$ varies from repulsive to attractive interactions, $b\in\{-0.75, 0.00, 2.75, 4.00\}$, where we set $r=0.2$ in (a) and $r=0.5$ in (b). Numerical data are obtained in the system of $L=2^{12}$ with the SB that is located at the bond of $(\frac{L}{2},\frac{L}{2}+1)$ (the middle of the system) and averaging over $10^8$ samples with $10^4$ configurations and $10^4$ different times, in the steady-state limit $(t\gg 2L^{3/2})$.}
\end{figure}

In Fig.~\ref{fig:Density}, we present density-profile patterns for various phases, where (a) $r=0.2$ (NS/PC)  and (b) $r=0.5$ (S/C). As $b$ varies from negative to positive, the phase of the system is changed from the NS/S to the PC/C, respectively. This corresponds to four cases in the middle and bottom panels of Fig.~\ref{fig:ConfigProfile}.

Based on the analytic results of the SB-free ZRP and the original SB problem with physical arguments, we develop MF approximations for the modified SB problem in the following subsection. Such approximations are valid if the interaction range is short enough to ignore correlations of the system. As long as a two-bulk picture is valid in the strong SB regime ($r\ll 1$), two-particle correlations can be ignored. 

%Pair MF approximations may go one step further than the single-site MF ones; however, they are invalid again as expected when the correlation length diverges in the maximal-current phase. Therefore, one needs to keep making longer cluster-MF approximations as much as possible for better results, but at the same time, they get more difficult to be solved. 

%

\subsection*{Mean-field (MF) approach} 
\label{sec:MF}

In this subsection, we focus on MF approximations to find the guidelines of phase boundaries in Fig.~\ref{fig:PhaseDiagram} (a). They are based on physical arguments for current-density relations and density-profile patterns.

\begin{figure}[]
	%Figure 5 : Density profiles (HD-LD Asymmetry, HD/LD coexistence diagram)
	\centering
	\includegraphics[width=\columnwidth]{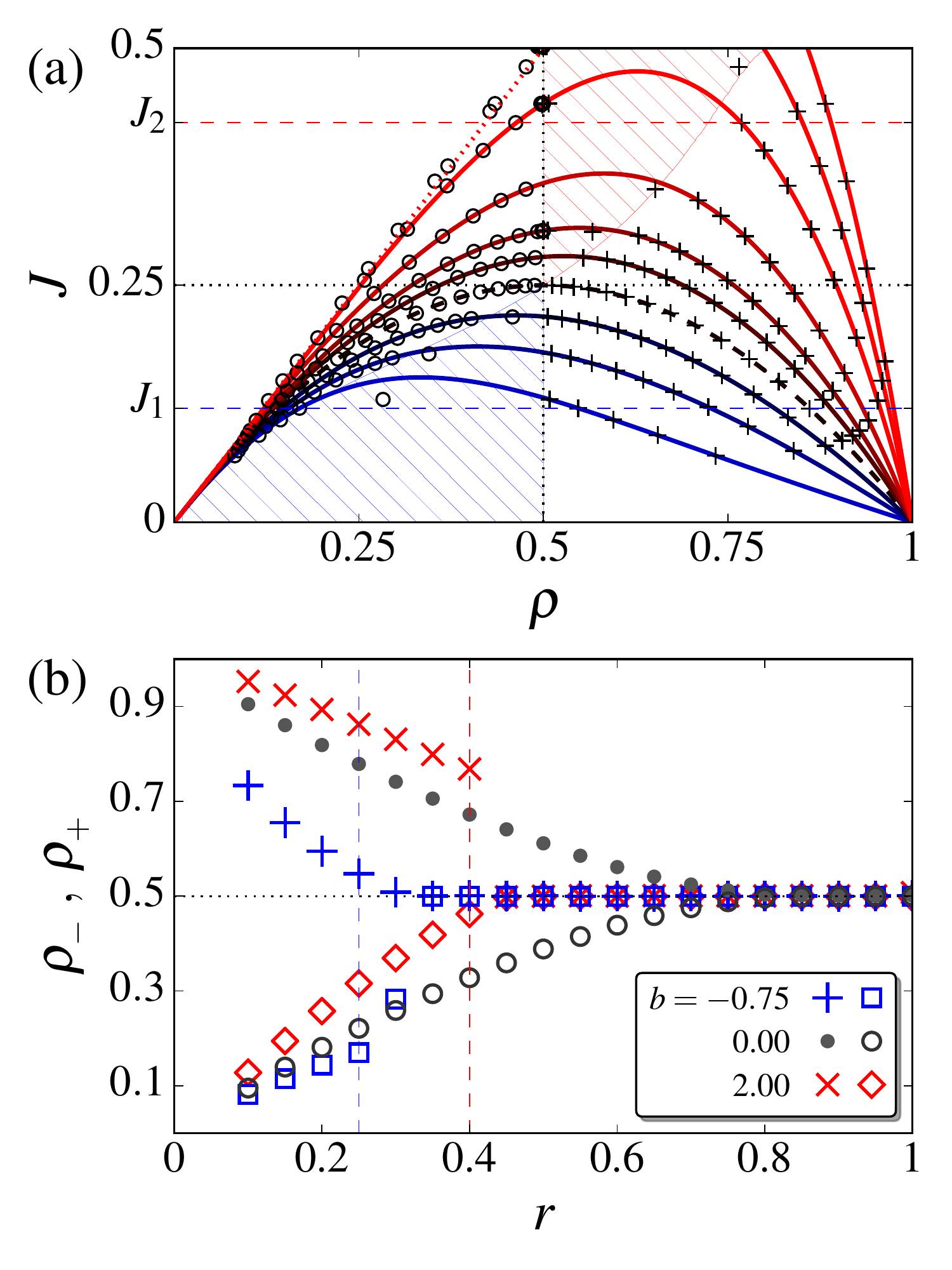}
	\caption{\label{fig:CoEx} For various $b \in \{$-0.75, -0.5, -0.25, 0, 0.25, 0.5, 1, 2, 3, 4$ \} $ (from blue to red) that are drawn as different colors, (a) the fundamental diagram by MF approximations is presented with numerical results. Numerically obtained high or low-density $\rho_{_{\pm}}$ (or $\rho=1/2$ in the NS phase) is plotted with different symbols $(+/\circ)$ for various $r$. Solid lines are drawn by Eq.~\eqref{eq:CurrByAvgRate} as $b$ varies, and the dashed line represents the ordinary TASEP ($b=0$). The forbidden-density regions (see the text for detailed discussion) are shown with shaded patterns. (b) Possible high and low densities are plotted as $r$ varies: the case of $b=-0.75$ and the case of $b=2.00$. The dashed guidelines are shown: ($b=-0.75, r=0.25$) with $J = J_1$ and ($b=2.00, r=0.40$) with $J = J_2$, respectively. It is noted that numerical data are overlapped as $r$ gets larger.}
\end{figure}

\begin{figure*}
	%Figure 6 : Current diagrams (current vs r, b)
	\centering
	\includegraphics[width=0.95\textwidth]{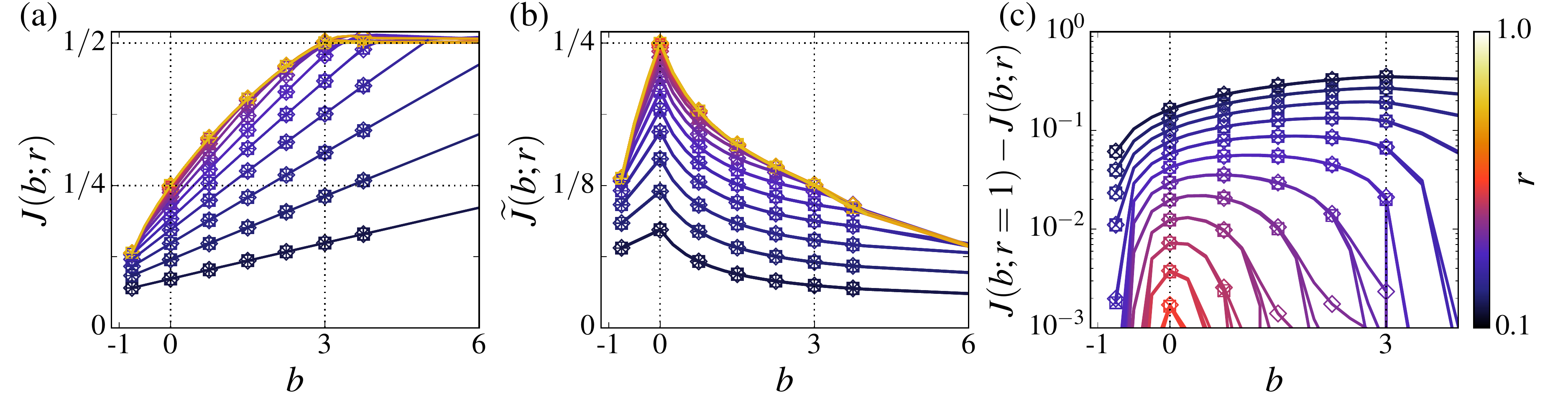}
	\caption{\label{fig:Currents} The average current $J(b; r)$ and related quantities are shown as a function of $b$. Each line represents the different values of $r$. The color scale from bright to dark is from $r=1$ to $r=0.1$: (a) the ZRP-type current $J$ with $J_{_{\rm max}}=\rho_{_0}=1/2$, (b) the conventional TASEP current $\widetilde{J}(\equiv J/u_{_{\rm max}})$ with the maximal current $\widetilde{J}_{_{\rm mc}}=1/4$, and (c) the relative-current difference between the current without the SB and that with the SB. Symbols represent different system sizes: $L=2^{10} (\diamond), 2^{12} (\times), 2^{14} (\square), 2^{16} (+)$. In (c), the splitting lines in the small value of current difference are caused by finite size effects at the NS-S boundary (see Sec.~\ref{sec:numerics} for detailed discussion).}
\end{figure*}

For the original SB problem, such MF treatments can be exact in the system of a single site with two particle reservoirs of $\rho_{_+}$ (left) and $\rho_{_-}$ (right), respectively, where the hopping rate between two is controlled by the SB factor $r$. Based on the current conservation, the local current has to satisfy the following relation: 
\begin{align}
J_{_{\rm MF}}=\rho_{_+}(1-\rho_{_+})=\rho_{_-}(1-\rho_{_-})=r\rho_{_+}(1-\rho_{_-}),
\label{eq:MF-current}
\end{align}
so that $\rho_{_-}=r\rho_{_+}$. When the higher-order correlations are regarded in the system with more sites, $\rho_{_-}/\rho_{_+} = r^* < r$ because higher-order terms decrease density separation to maximize the global current of the system. Therefore, $r^*$ acts as the upper limit of the density ratios. We use this relation, together with Eq.~\eqref{eq:MFCriteria} and the current-density relation to find a functional form of $r^*(b)$ for the S-NS phase boundary. However, there are no closed-form expressions of the current-density relation for arbitrary values of $b$ and MF approximations of $J(\rho, b)$ are used just as the guideline of the phase boundaries. We compare them with numerical results (see Figs.~\ref{fig:CoEx} and~\ref{fig:Currents}). 

The current of the system under the influence of the small SB factor increases linearly with $b$. This is due to the current of high-density parts that is usually influenced by a single-site hopping rate, and leads to that the current is linearly proportional to $b$; see Fig.~\ref{fig:Currents} (a). Based on numerical observations and physical arguments, we estimate the current $J(b;r)$ around the limit of $|b| \rightarrow 0$. Using the MF equation of the original SB problem, the current and the high (low) bulk density can be expressed by the expansion up to the first order of $b$: 
\begin{align*}
J(b;r)= \frac{r}{(1+r)^2} + b g(r),\\
\rho_{_+}= \frac{1}{1+r} +b f_{_+}(r);~\rho_{_-} = \frac{r}{1+r} + b f_{_-}(r).
\end{align*}
%where $g, f_{_+}, f_{_-}$ are perturbative term in presence of SB.

The MF current of the modified TASEP is simply 
\begin{align}
J_{_{\rm MF}} = \rho \phi(\rho),
\label{eq:CurrByAvgRate}
\end{align}
where $\phi(\rho)=\langle u(\ell_x) \rangle$ (see Appendix~\ref{appendix2} for details).
The average hopping rate (the phase velocity) $\phi_{_\pm} = J/\rho_{_\pm}$ also has asymmetry in the presence of $b$. The current-density relation up to the first order of $b$ (see Eq.~\eqref{eq:phi-rho-neutral} in Appendix~\ref{appendix2}),
\begin{align*}
J_{_{\rm NS}}(b;r=1) = \frac{\rho(1-\rho)}{1-b\rho},
\end{align*}
which is drawn in Fig.~\ref{fig:CoEx} (a).
Applying the steady-state current conservation across the SB, 
\begin{align*}
\rho_{_+}\phi_{_+} = \rho_{_-}\phi_{_-} = r\rho_{_+}\phi_{_-}.
\end{align*}
As a result, we estimate the results of Figs.~\ref{fig:CoEx} and~\ref{fig:Currents} (a) as follows:
\begin{align} 
\label{eq:J_r_b}
J_{_{\rm S}}(b;r) & = \frac{r}{(1+r)^2} + b \frac{2r^2}{(1+r)^4},
\end{align}
and 
\begin{align} 
\label{eq:rho-p-0}
\rho_{_+} & = \frac{1}{1+r} \left\{ 1 + b \frac{r}{(1+r)^2}\right\},\\
\label{eq:rho-m-0}
\rho_{_-} & = \frac{r}{1+r} \left\{ 1 + b \frac{r}{(1+r)^2}\right\}.
\end{align}

Using the conservation of the current, we estimate the boundary between density-separated phases (both S and PC) and the uniform density phases (both NS and C). As $b$ gets larger, the SB effect becomes weaker and weaker, so that the difference between the current without the SB and that with the SB gets smaller and smaller. Eventually, the SB effect is completely suppressed as if $r=1$ due to the role of $b$. The criterion leads $r^*(b)$ far from $b=0$:
\begin{align} 
\label{eq:PB-NS-S}
J_{_{\rm S}}(b;r^*(b)) = J_{_{\rm NS}}(b;1),
\end{align}
where $J(b;1)$ is the current of the modified TASEP without the SB. Since the closed-form expression of Eq.~\eqref{eq:PB-NS-S} doesn't exist, we draw its numerical solutions as different colored (red) lines for $0<b<3$ in Fig.~\ref{fig:PhaseDiagram} (a).
Similarly, when $J(b>3;1) =J_{_{\rm max}}$ and $J_{_{\rm max}}=\rho_{_0}=1/2$, together with Eq.~\eqref{eq:J_r_b}, the C-PC phase boundary is shaped as 
\begin{align} 
\label{eq:PB-C-PC}
b(r) = \frac{(1+r)^2(1+r^2)}{4r^2}.
\end{align}
Moreover, the PC phase can be distinguished from the S phase. Since we observe the partial condensate of holes only in the low-density part of the PC phase, so $\phi_{_-}=1$. As a result,
\begin{align}
\label{eq:J-PC} 
J_{_{\rm PC}} = \rho_{_-} = \frac{b-2}{b-1},
\end{align}
which can be obtained from the special case of Eq.~\eqref{eq:rho-F} in Appendix~\ref{appendix2}. Therefore, at the S-PC phase boundary, Eq.~\eqref{eq:J-PC} is equal to Eq.~\eqref{eq:J_r_b}, which is implicitly expressed as
\begin{align} 
\label{eq:PB-S-PC}
\frac{b_{_{\rm S/PC}}-2}{b_{_{\rm S/PC}}-1} = \frac{r}{(1+r)^2} + b_{_{\rm S/PC}} \frac{2r^2}{(1+r)^4}.
\end{align}

As $b$ is larger than the critical value $b_{_{\rm C}}$ for a given value of $r$, vacancies (holes) are condensed and formed as a macroscopic cluster, namely, the full condensation. The critical value $b_c$ depends on the density of the system when the number of particles is conserved, which was calculated in the ZRP study by Grosskinsky {\it et al.}~\cite{Grosskinsky2003} (see Appendix~\ref{appendix2} for details).
\begin{align}
\label{eq:PB-NS-C}
\frac{1}{b_{_{\rm C}}-2} = \langle \ell \rangle = \frac{1-\rho_{_0}}{\rho_{_0}}.
\end{align}
It implies that the ordinary ZRP condensates at $b_{_{\rm C}}=3$ for $\rho_{_{\rm C}}=\rho_{_0}=1/2$. Using this criterion of Eq.~\eqref{eq:PB-NS-C}, we find the NS-C phase boundary at $b_{_{\rm C}}=3$, which works well as long as the SB effect is weak enough to be ignored. As $r$ gets smaller, it should be compared to the criterion of the PC phase, 
where $\phi_{_-}=1$ and $\rho_{_{\rm PC}}=\rho{_{-}}(r)<\rho_{_0}$, so that we get the C-PC phase boundary as: 
\begin{align}
\label{eq:PB-PC/C}
\frac{1}{b_{_{\rm PC/C}}-2}=\frac{1-\rho_{_-}}{\rho_{_-}},
\end{align}
where $\rho_{_-}$ is obtained from Eq.~\eqref{eq:rho-m-0}.

When $r \rightarrow 0$, $\rho_{_-}\rightarrow 0$ leads to $b_{_{\rm PC}} \rightarrow 2$, denoting the left endpoint of the S-PC phase boundary at $r=0$ and $b=2$. On the other hand, the right endpoint of the S-PC phase boundary is obtained when the $J_{_{\rm PC}} = \rho_{_{\rm PC}}=\rho_{_-}$ and $\rho_{_-}=\rho_{_0}=1/2$, which coincides with the PC-C phase boundary at $b=3$. Moreover, the location of $r_{_{\rm S/PC/C}}$ is the specific value from the solution of $J(r, 3)=1/2$, so that it ends at $r_{_{\rm S/PC/C}}\approx 0.475$ and $b=3$.

In the neutral ($|b| \rightarrow 0$) regime, we are able to use the marginal density relation Eq.~\eqref{eq:MFCriteria} and the current-density relation up to the first order $b$ [see Eq.~\eqref{eq:phi-rho-neutral}]. From Eq.~\eqref{eq:rho-p-0} and Eq.~\eqref{eq:rho-m-0}, we observe $\rho_{_-}=r\rho_{_+}$. This is the limit when all correlations other than sites next to the SB are neglected. The resulting MF approximations around $r=1$ and $b=0$ provide both NS-S and S-NS phase boundaries as:
\begin{align} 
\label{eq:PB-b0}
b_{_\pm}(r) =2\left( r^{\mp 1}-1 \right)
\end{align}
where the sign corresponds to the sign of $b$, so $b_{-}$ is the NS-S phase boundary for $b<0$ and $b_{+}$ is the S-NS phase boundary for $b>0$. We draw these NS-S-NS boundaries as different colored (green, solid and dotted) lines up to $|b|<1$ in Fig.~\ref{fig:PhaseDiagram} (a). 

In the next section, we present extensive Monte-Carlo (MC) simulation results, compared with MF predictions that have been discussed so far, where we explain all the figures and some interesting features as well as some discrepancies between numerical results and MF ones. 

\section{Numerical Results}
\label{sec:numerics}

\begin{figure*}[]
    %Figure 7 : Condensation
	\centering
	\includegraphics[width=\textwidth]{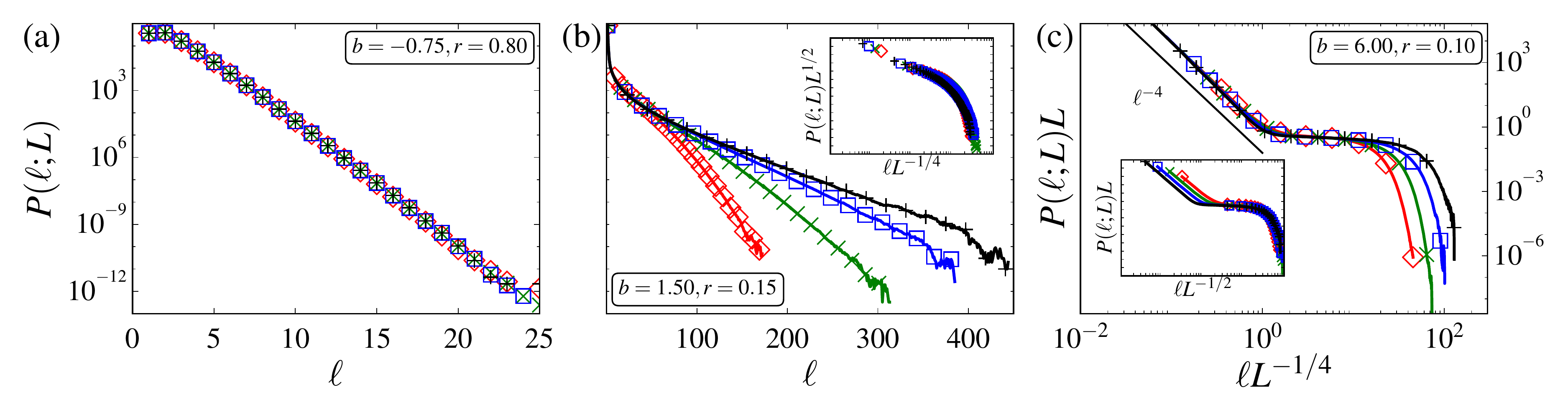}
	\caption{\label{fig:Condensation} The distribution of inter-particle distance $P(\ell;L)$ is plotted for $L=2^{10} (\diamond), 2^{12} (\times), 2^{14} (\square), 2^{16} (+)$ at three different phases: (a) The NS phase 
	at $(b=0.75, r=0.80)$ shows $P(\ell;L)\sim \exp(-\ell/\ell_{_{\rm NS}})$, where $\ell_{_{\rm NS}}$ is independent of $L$; (b) the S phase at (1.5, 0.15) shows $P(\ell;L)\sim L^{-1/2}\exp(-\ell/L^{1/4})$; (c) the PC phase at (6.00, 0.10) shows $P(\ell;L)L\sim f_{_{\rm PC}}(x_{1}, x_{2})$, where $f_{_{\rm PC,1}}(x_1)=x_{1}^{-4}$ for $x_{1}< 1$, $f_{_{\rm pc,1-2}}(x)=\mbox{constant}$ for $x_1<x<x_2$, and $f_{pc,2}(x_2)\sim\exp(-x_2)$ for $x_2> 1$ with $x_{1}\equiv\ell/L^{1/4}$ and $x_{2}\equiv\ell/L^{1/2}$. In particular, scaling collapses are tested in the inset of (b) and (c).}
\end{figure*}

%%Numerical method
Our numerical data are taken from the modified TASEP with a SB for various system sizes of $L\in\{2^{10}, 2^{12}, 2^{14}, 2^{16}\}$. The SB is located in the middle of the system at the bond $(\frac{L}{2},\frac{L}{2}+1)$ with the SB factor $r$. Initially, the system is prepared with the alternative particle-hole configuration for $\rho_{_0}=1/2$. The system is relaxed to reach the stationary state after $t=2L^{3/2}$ MC steps from the initial condition.

Figure~\ref{fig:Density} shows typical patterns of density profiles. The fundamental diagram where numerical data are shown as symbols is given in Fig.~\ref{fig:CoEx}, and MF predictions as lines. To distinguish four phases accurately, we measure both the ZRP-type current $J$ and the conventional TASEP current $\widetilde{J}$ as a function of $b$ for various $r$ in Fig.~\ref{fig:Currents}. 

%Once the system is stabilized, particle occupation is added by each site, per every $L^{3/2}$ steps, $10^6$ to $10^3$ step (for increasing system size) in total. Throughout measurement, number of hopping events are also summed and divided into total steps to measure current.

%Abount density and Measurement of the homogeneous bulk density
The jamming of particles caused by the SB can be directly observed from density profiles. In the strong SB regime, the density behind the SB contains the nonvanishing extra density rather than the average global density as the macroscopic queue (S phase), while in the relatively weak SB regime, there is no extra density (NS phase). 

However, it is a difficult task to precisely measure the bulk density from density profiles~\footnote{Previous studies~\cite{MHa2003,HSoh2017} used the average particle occupation at $x=L/4$ as a good proxy of the bulk density, since the particle-hole symmetry in the original TASEP manifests the boundary of HD/LD locates at $x=L/2$. With the broken symmetry ($b \neq 0$), the boundary does not locate at $x=L/4$.}. Using the distribution of the average occupation per site, $P(\langle n_x\rangle)$, we measure the density difference $\Delta = \rho_{_+}-\rho_{_-}$. We assume $P(\langle n_x\rangle)$ to be a Gaussian as follows: In general, the functional form of $P(\langle n_x \rangle)$ is not rigorously proven as a Gaussian, but it is a reasonable assumption to find the location of the peak without loss of generality: 
\begin{align*}
P(\langle n_x\rangle) =
\begin{cases}
\mathcal{N}_{\rho, \sigma}(\langle n_x\rangle) & {\rm (NS)} \\
c_+ \mathcal{N}_{\rho_+, \sigma_+}(\langle n_x\rangle)+ c_-\mathcal{N}_{\rho_-, \sigma_-}(\langle n_x\rangle) & {\rm (S)} 
\end{cases}
\end{align*}
where $\mathcal{N}_{\mu, \sigma}$ is the normal distribution with the average $\mu$ and the standard deviation $\sigma$. 

In Fig.~\ref{fig:PhaseDiagram}, we also numerically provide a phase diagram. In Fig.~\ref{fig:ConfigProfile}, snapshots of typical spatiotemporal patterns are presented for various phases. In Fig.~\ref{fig:Density} at (a) $r=0.2$ and (b) $r=0.5$, we show typical patterns of density profiles, which are analyzed as $P(\langle n_x\rangle)$. Unlike the ordinary TASEP ($b=0$) where the excess bulk density is symmetric, the modified TASEP ($b\ne 0$) exhibits particle-hole asymmetry, because exchanging a particle as a hole $\rho \rightarrow (1-\rho)$ and the hopping direction $x \rightarrow -x$ does not reproduce the same result. In the separate phase, two bulks are separated by the SB and have different densities, while, in the NS phase, the SB effect is localized and the bulk is uniform: $\langle n_x\rangle \approx 1/2$. 

Using the results of $P(\langle n_x\rangle)$, we can identify the bulk density $\rho$. In Fig.~\ref{fig:CoEx} (a), we plot the current $J$ as a function of $\rho$, where both $J$ and $\rho$ are measured from MC simulations as well as the current-density relation by MF approximations for the homogeneous system with $b \in \{-0.75, -0.5, -0.25, 0, 0.25, 0.5, 1,2,3,4\}$ from bottom to top. At the same $b$, the system may be in the NS phase if $\rho_{_-}=\rho_{_+}=1/2$. As the SB effect becomes strong, the system is split into two subsystems with nonzero density separation. This process is shown for $b<0$ in (b) and $b>0$ in (c). Due to the total-density conservation, the high (low) density jumps suddenly when the density separation happens in $b>0$ ($b<0$), with the inaccessible gap between them. This gap is numerically obtained by using Eq.~\eqref{eq:MFCriteria}. For a detailed description of the current-density relation, see Appendix~\ref{appendix2}. In the presence of the SB from $r=0.1$ to $r=1$, we provide a specific example of $b=2.50$ in Fig.~\ref{fig:Apx_Coex}. Two bulks lie on the homogeneous current-density relation since separated bulks behave as independent homogeneous systems with the same current. When $2<b<3$, $\rho(\phi=1)=(b-2)/(b-1)$. When the $\rho_{_-}<(b-2)/(b-1)$, the $\rho_{_-}$ lies on the $\phi=1$ line, which denotes the partial condensation of holes in the low-density part.

In Fig.~\ref{fig:Currents}, we redraw the current as a function of $b$ for various $r$: (a) The ZRP-type current is maximized up to $\rho_{_0}=1/2$ and $b(r)$ can be found as the NS-C boundary if $r$ is large enough to see the flat region; (b) the conventional TASEP current is maximized up to $\tilde{J}_{_{\rm mc}}=\rho_{_0}(1-\rho_{_0})=1/4$ and the peak is located at $b=0$. Whether the SB is localized or not can be measured, in the context of the relative-current difference between the system without and with the SB, $J(b;r=1)-J(b;r)$. When the jamming of particles is globally expanded, the current difference is finite. In the other limit, the relative-current difference is strictly nonzero but inversely proportional to the system size as the SB effect is localized. The detailed values are shown in (c), where the relative-current difference converges to zero rapidly as $b\to -1$ and $b\gg 0$. It is noted that as in the smaller difference, the system size dependence comes in and the relative-current difference is shown by splitting lines by different symbols. This is analogous to the crossover scaling found in the ordinary SB problem~\cite{HSoh2017}.

Finally, we discuss the interesting scaling features of inter-particle distance distributions $P(\ell;L)$ shown in Fig.~\ref{fig:Condensation}, where three different phases are compared with the characteristic length $\xi$ for various system sizes: (a) the NS phase at $(b=-0.75, r=0.8)$, (b) the S phase at (1.50, 0.15), and (c) the PC phase at (6.00, 0.10). As $b$ increases but still for $b<b_{_{\rm S/PC/C}}$, $\xi$ gets longer but is independent of $L$. However, passing the NS-S phase boundary, it becomes a power law as a function of $L$, \textit{e.g.}, $\xi\sim L^{1/4}$. Even further, another length scale comes in as the size of the partial condensate in the low-density part, $\ell_{_{\rm PC}}\sim L^{1/2}$. 

Scaling collapses are also tested in Fig.~\ref{fig:Condensation} as the inset of panels (b) and (c). 
In the NS phase,
\begin{align}	
P(\ell;L)L^{1/2}=f_{_{\rm NS}}(\ell/\xi_{_{\rm NS}}),
\end{align}
where $\xi_{_{\rm NS}}\sim \mbox{constant}$, depending on $b$ and $r$ only, and $f_{_{\rm NS}}(x)\sim\exp(-x/\xi{_{\rm NS}})$.
In the S phase,
\begin{align}
P(\ell;L)L^{1/2}=f_{_{\rm S}}(\ell/L^{1/4}),
\end{align}
where $\xi_{_{\rm S}}\sim L^{1/4}$, depending on $b$ and $r$ as well, and $f_{_{\rm S}}(x)\sim\exp(-x/\xi{_{\rm S}})$. In the PC phase,
\begin{align}
P(\ell;L)L=
	\begin{cases}
	f_{_{\rm PC,1}}(\ell/L^{1/4}), &\mbox{(normal)};\\ 
	f_{_{\rm PC,2}}(\ell/L^{1/2}), &\mbox{(PC)}.
	\end{cases}
\end{align}
 
Approaching from the NS phase to the S phase, the characteristic interparticle distance becomes longer but still finite as a constant independent of $L$. Passing the NS-S phase boundary, it eventually depends on $L$ and follows specific power-law scaling in the S phase. Moreover, in the PC phase, the condensate of holes develops, which scales as $\ell\sim L^{1/4}$ up to $\ell\sim L^{1/2}$ in the low-density part. Up to $\ell <L^{1/4}$, $P(\ell;L)\sim \ell^{-4}$, independent of $b$ as long as the system in the PC phase. The origin of scaling in the large $\ell$ regime mostly depends on the low-density part, as in the high-density part mostly contributes to the small $\ell$ regime.

For the low-density part in the PC phase, particles randomly inject and have the geometric distribution of $\ell$, very near the SB. Then, as particles travel forward, vacancies (holes) form a cluster from the random initial cluster. Even this process is in the stationary state, the condensation process along the spatial axis is equivalent to the dynamic cluster formation of the ordinary ZRP. Therefore, in the condensate region, the phase velocity is equal to unity; the spatial position $x$ is equivalent to the coarsening time $t$ from the random initial condition (see Fig.~\ref{fig:ConfigProfile} at $b=4.00$ and $r=0.2$). As a result, the inter-particle distance distribution in the low-density part is the same as the integrated cluster-size distribution from $t = 0$ to $t = cL$, where $c$ is the fraction of the low-density bulk ($1/2 <c<1$). Therefore, that average cluster size for the totally asymmetric ZRP scales as $\langle \ell\rangle_t \sim t^{1/2}\sim L^{1/2}$~\cite{Grosskinsky2003}, where the number of the condensed cluster is the order of unity, and the time scales as $L$, leading to $\ell_{_{\rm PC, C}}\sim L^{1/2}$. Additional figures are available as Fig.~\ref{fig:Condensation-test} in Appendix~\ref{appendix3} for the comparison of the ZRP-type condensation in the C phase with other phases.
 
\section{Summary and Discussion}
\label{sec:summary}

%Summary of the work
We have studied the interplay of particle interaction and local defect in the current-density relation under the conservation of particles and the global current through the entire system. In our study, we considered the modified slow-bond (SB) problem with two well-known nonequilibrium models, the totally asymmetric simple exclusion process (TASEP) and the zero-range process (ZRP). In the modified SB problem, the interaction parameter $b$ and the SB factor $r$ are two main control parameters. 

As $b\in(0,\infty)$ and $r\in(0,1]$ vary, the phase diagram was suggested with marginal phase boundaries that are obtained from mean-field (MF) approximations for the SB problem, which were also numerically checked. In particular, we found that the particle-hole asymmetry due to the ZRP-type hopping rates allows the system to have the nonseparated (NS) density profiles, which is called the NS phase, {\it i.e.}, the SB-free phase. As a result, in the modified SB problem, jamming caused by the SB can be localized in the thermodynamic limit, which is different from the original SB problem. However, in this paper, finding exact boundaries of the SB-free phase and the scaling relations near transitions is out of our scope, which will be discussed elsewhere as one of our future studies. 

On the other hand, the modified SB problem would shed light on similar issues in real-world traffic and transport problems, such as a localized blockage in highways and metabolic systems. The correlated dynamics in driven flow is closely related to the interparticle distance-dependent hopping rate. Real-world traffic and transport problems are often treated as cellular automata and biased random-walk type models with correlated physical quantities. This is quite similar to $b \leq 0$, where our results imply that particle interaction can suppress the jamming of particles. Moreover, it would be interesting to test the rich and robust scaling behaviors of the partially condensed phase, obtained from the inter-particle distance distribution function in the open system as well, which would be another challenging task, in the context of the ensemble equivalence.

\section*{Acknowledgments}
This research was supported by the Basic Science Research Program through the National Research Foundation of Korea (NRF) (KR) NRF-2017R1D1A3A03000578 (H.S., M.H.), NRF-2012-SIA3A-2033860 (H.J.), and also by a research fund from Chosun University (KR), 2012 (M.H.).

\appendix 
\counterwithin{figure}{section}

\section{Mapping of modified TASEP onto ZRP} 
\label{appendix1}

\begin{figure}[b]
	%Figure A1 : Current-density relation in homogeneous system and SB explanation.
	\centering
	\includegraphics[width=\columnwidth]{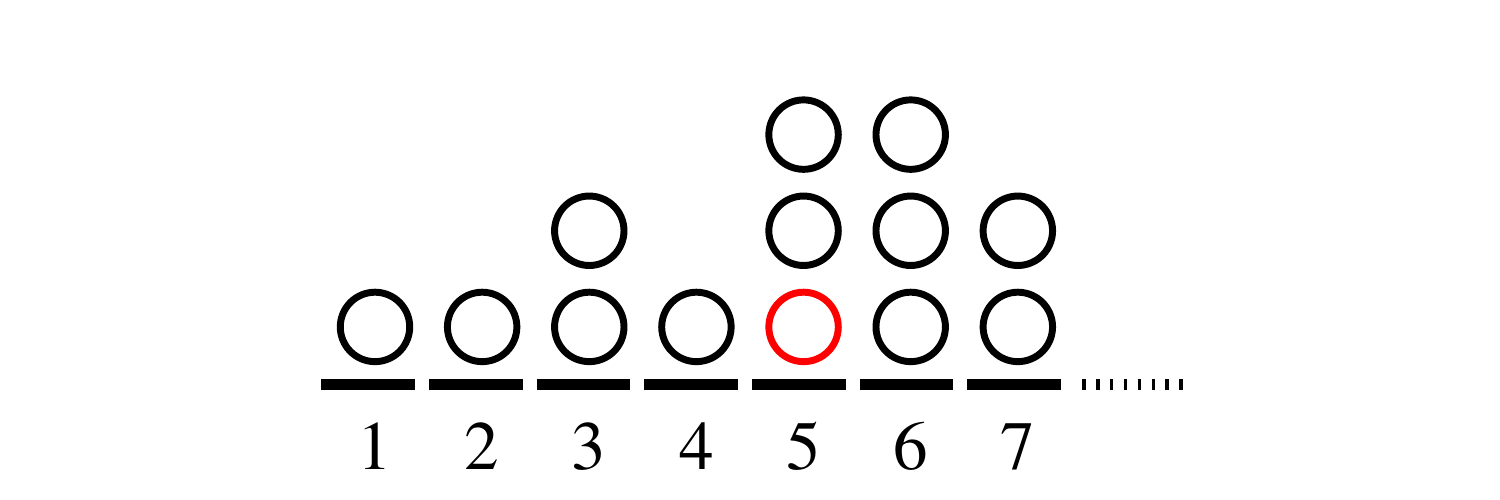}	
	\caption{\label{fig:mapping-zrp} The ZRP-type mapping of Fig.~\ref{fig:model} is illustrated, where the site index with number represents the particle index in the modified TASEP, and particles at each site represent links between consecutive particles in the modified TASEP. The slow bond becomes a slow particle as drawn in the different colored (red) circle.} 
\end{figure}

The standard mapping of the TASEP onto the ZRP is exact only if there is no defect and the system is periodic. The mapping is a unique one-to-one correspondence between particles (empty sites in front of the chosen particle) in the TASEP and sites (particles at the mapped site) in the ZRP. In the modified TASEP, such mappings are no longer exact due to the presence of the SB. 

However, there is one possible remedy as shown in Fig.~\ref{fig:mapping-zrp}: Particles in the modified TASEP still are mapped to sites in the ZRP-type dynamics, while particles in the ZRP-type dynamics are not the number of empty sites between two consecutive particles in the modified TASEP but the number of links between them. The SB can be treated as a slow particle in the ZRP-type dynamics. To map the modified TASEP dynamics onto in the ZRP-type one, the first arrival particle among at least two particles at the chosen site, can move only to the neighboring site.    

\section{Current-Density Relation\\ for Homogeneous Case} 
\label{appendix2}

In the TASEP, the system with $N$ particles and $(N+L)$ sites corresponds to the ZRP with $L$ particles and $N$ sites. This can be described as the stationary process of $\{\ell_x\}$:
\begin{align}
P^{N, L}\{\ell_x\} = \frac{1}{Z(N, L)} 
\prod_{x=1}^{N} W(\ell_x)\delta \left(N, \sum_L\{\ell_x\} \right),
\end{align}
where the weight $W$ is given by
\begin{align}
W(\ell)=\prod_{i=1}^{\ell} \frac{1}{u(i)},
\end{align}
and the normalizing partition function $Z$ is given by
\begin{align}
Z(N,L)=\sum_{\{\ell_i\}} \prod_{x=1}^{L}W\{\ell_x\}\delta \left(N, \sum_L\{\ell_x\}\right).
\end{align}
The equivalence of canonical and grand-canonical ensembles~\cite{Grosskinsky2003} defines the grand-canonical measure as:
\begin{align}
P^L_\phi \{\ell_i\} = \prod_{x=1}^{L} P_\phi(\ell_x),
\end{align}
with the single-site measure and its normalization
\begin{align}
P_\phi(\ell_x) = \frac{1}{\mathcal{Z}}W(\ell_x) \phi^{\ell_x}, \\
\mathcal{Z}(\phi) = \sum_{\ell_x = 0}^{\infty} W(\ell_x) \phi^{\ell_x}.
\end{align}
In the grand canonical ensemble, the average particle density $\langle \ell \rangle (\phi)$ as a function of $\phi$ is given by
\begin{align}
\langle \ell \rangle (\phi) = \sum_{n_x=0}^{\infty} \ell_x P_\phi(\ell_x) = \phi \frac{\partial}{\partial \phi} \ln \mathcal{Z}.
\end{align}
The average velocity (jump rate) is the expectation value of hopping rates:
\begin{align}
\langle u(\ell_x) \rangle = \sum_{\ell_x = 0}^\infty u(\ell_x) P_\phi (\ell_x) = \phi.
\end{align}
As a result, for $u(\ell_x)$ studied in~Ref.\cite{Evans2000}, the stationary weight for the process with $b$ is given by
\begin{align}
W(k) = \prod_{i=1}^{k} \frac{1}{1+b/i} = \frac{\Gamma(k+1) \Gamma(1+b)}{\Gamma(1+b+k)}.
\end{align}
The grand-canonical partition function can be written in terms of the hypergeometric function~\footnote{M. Abramowitz. Handbook of Mathematical Functions. Dover, New York, 1972.},
\begin{align}
\mathcal{Z} = {}_2F_1(1,1;1+b;\phi)=\sum_{k=0}^{\infty} \frac{\Gamma(k+1) \Gamma(1+b)}{\Gamma(1+b+k)} \phi^k,
\end{align}
as well as the average particle distance 
\begin{align}
\langle \ell \rangle (\phi) = \frac{\phi}{(1+b)} \frac{{}_2F_1(2,2;2+b;\phi)}{{}_2F_1(1,1;1+b;\phi)},
\end{align}
which leads to $\rho$ as,
\begin{align}
\label{eq:rho-F}
\rho = \frac{1}{1 + \langle \ell \rangle} = \frac{{}_2F_1(1,1;1+b;\phi)}{{}_2F_1(1,2;1+b;\phi)},
\end{align}
where the latter relation is from the hypergeometric identity. The value of the hypergeometric function for $\phi=1$ when $c-a-b > 0$ is
\begin{align}
{}_2F_1(a,b;c;1) = \frac{\Gamma(c)\Gamma(c-a-b)}{\Gamma(c-a)\Gamma(c-b)}.
\end{align}
This gives us $\rho$ at given $b$ with $\phi=1$ as
\begin{align}
\rho(1) = \frac{b-2}{b-1}.
\end{align}

%In the presence of the SB (a slow particle in the ZRP), the system is split into two subsystems with density separation around the SB. In the stationary and thermodynamic limit, the density-separated (S) phase consists of two homogeneous subsystems with the equal particle flux. This corresponds to the traveling high-density $\rho_+$ packet in a pool of low-densities in the ZRP picture. In this case, the average speed of the packet is equal to the average speed of the slow particle. 
%

\begin{figure}[b]
	%Figure B1 : Current-density relation in homogeneous system and SB explanation.
	\centering
	\includegraphics[width=\columnwidth]{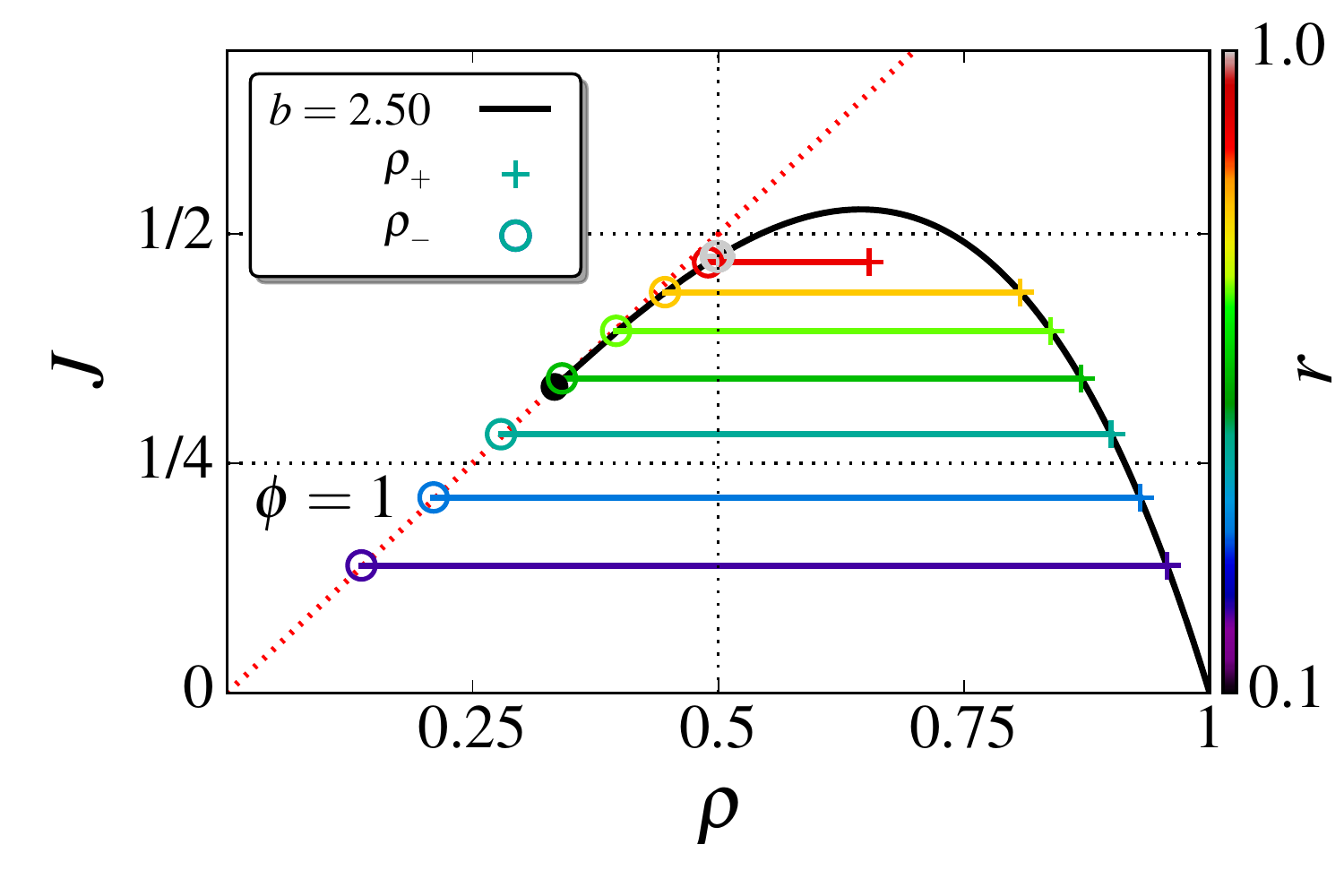}	
	\caption{\label{fig:Apx_Coex} An example of the current-density relation.} 
\end{figure}

\subsection*{Neutral limit: $|b| \to 0$}

In the case of small $b$, the current-density relation can be obtained from the perturbation of the partition function. Using the Euler hypergeometric transformation, the partition function $\mathcal{Z}$ is expanded in terms of $b$,

\begin{align} 
\label{eq:rho-phi}
\mathcal{Z} &= {}_2F_1(1,1;1+b;\phi) = \frac{{}_2F_1(1,b;1+b;\frac{\phi}{\phi-1})}{1-\phi}, \nonumber\\
&= \frac{1-\sum^\infty_{k=1} (-b)^k{\rm Li}_k(\frac{\phi}{\phi-1})}{1-\phi},
\end{align}
where ${\rm Li}_k(z)$ is the polylogarithmic function. Using the polylogarithmic identity,
$\rho$ is expressed in simple terms:
\begin{align}
\rho(\phi) &= \frac{1}{1+\langle \ell \rangle} = \frac{1-\phi}{1+b \left(\frac{1}{\mathcal{Z}}-1\right)}.
\end{align}
By the series inversion, we get $\phi(\rho)$ up to the several orders of $b$, 
\begin{align}
\phi &= (1-\rho) + b \rho (1-\rho) + b^2 \rho^2 \left[ (1-\rho)+ \ln \rho \right] 
\nonumber\\
&+ b^3 \rho^2 \left[{\rm Li}_2(1-\rho)+(2\rho-1)\ln \rho -\frac{1}{2}\ln^2\rho-(1-\rho)^2 \right]\nonumber\\
&+ \mathcal{O}(b^4),
\end{align}
as well as the current $J = \rho \phi(\rho)$, and we retrieve the original TASEP current $\rho (1-\rho)$ as $b \to 0$. This expansion does not have the closed form, we approximate up to the first order of $b$ in Eq.~\eqref{eq:rho-phi}. As a result,
\begin{align}
\label{eq:phi-rho-neutral}
\phi = \frac{1-\rho}{1-b\rho}.
\end{align}

In Fig.~\ref{fig:Apx_Coex}, we present the fundamental diagram for $J(\rho; b=2.50)$ as well as $\rho_{_+}$ and $\rho_{_-}$ for various $r$ from $r=0.1$ (violet) to $r=1.0$ (gray). The black solid line is drawn by Eq.~\eqref{eq:CurrByAvgRate} for $b=2.50$, where the left-side endpoint coincides with $\phi=1$ as shown the different colored (red, dotted) line. As $r$ decreases, $\rho=1/2$ (NS), $\rho_{_+}$ and $\rho_{_-}$ are also marked.

\begin{widetext}
\section{Condensation Analysis} 
\label{appendix3}

Figure~\ref{fig:Condensation-test} presents the additional information for the comparison of the ZRP-type condensation in the C phase with other phases as $P(\ell;L=2^{16})$: In the upper panel at $r=0.10$ (left), $r=0.40$ (middle) and $r=0.95$ (right) as $b$ varies from $b=-0.75$ (black) to $b=6.00$ (yellow), we find that $P_{_{\rm C}}(\ell)\sim \ell^{-b}$ for $b\ge 3$ in the C phase, while, in the PC phase, $P_{_{\rm PC}}(\ell)\sim \ell^{-4}$ for $b\gg b_{_{\rm PC}}(r)$. In the lower panel at $b=2.50$ (left), $b=3.00$ (middle), and $b=6.00$ (right) as $r$ varies from $r=0.10$ (black) to $r=0.95$ (yellow), we confirm that the functional shape of $P(\ell)$ corresponds to the phase identity. The guidelines of slopes are provided: In the upper panel, the long-dashed lines are -4.0 (black), -3.0 (green), and -6.0 (blue), respectively. In the lower panel, the black long-dashed lines are -4.0, while the red long-dashed lines correspond to $-b$.

\begin{figure*}[h]
\centering
	\includegraphics[width=0.95\textwidth]{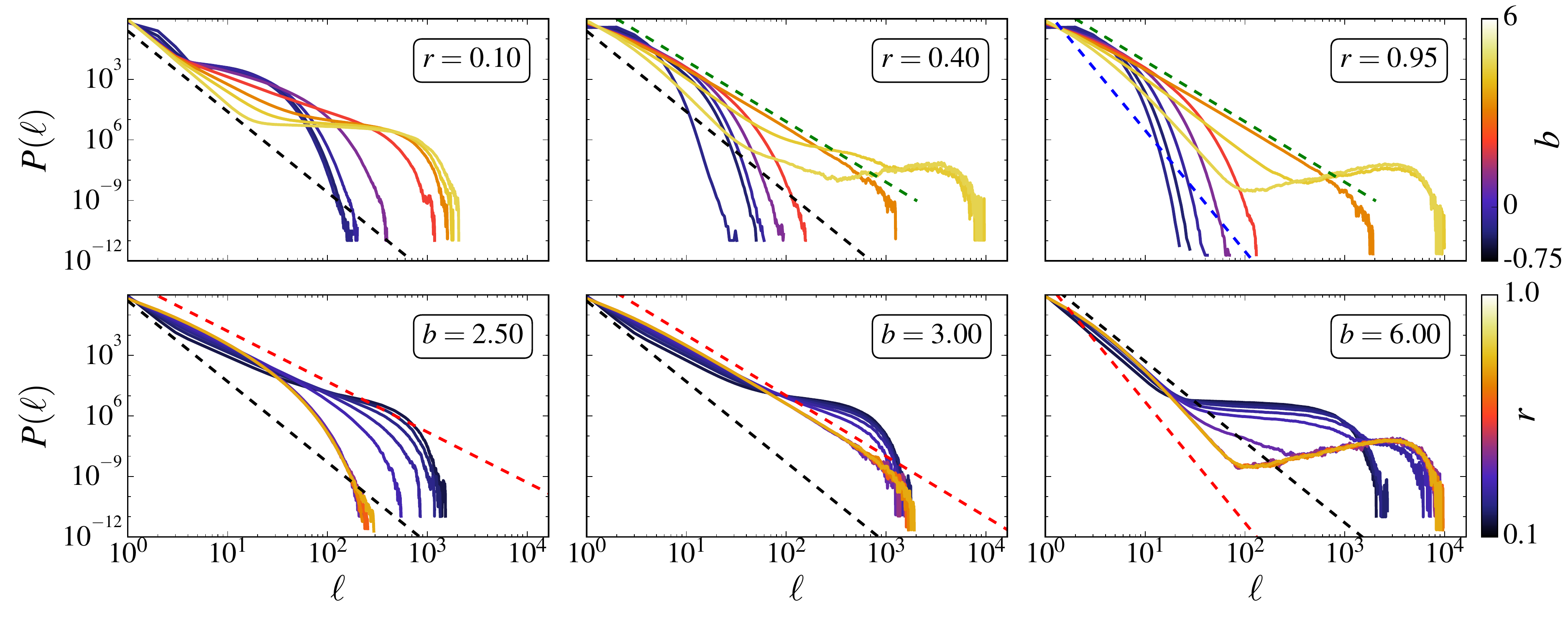}
	\caption{\label{fig:Condensation-test}
	Inter-particle distance distributions for $L=2^{16}$ for 6 settings.}
\end{figure*}
\end{widetext}

\bibliography{ref-PRE2017-SHJ}

\begin{thebibliography}{33}%
\makeatletter
\providecommand \@ifxundefined [1]{%
 \@ifx{#1\undefined}
}%
\providecommand \@ifnum [1]{%
 \ifnum #1\expandafter \@firstoftwo
 \else \expandafter \@secondoftwo
 \fi
}%
\providecommand \@ifx [1]{%
 \ifx #1\expandafter \@firstoftwo
 \else \expandafter \@secondoftwo
 \fi
}%
\providecommand \natexlab [1]{#1}%
\providecommand \enquote  [1]{``#1''}%
\providecommand \bibnamefont  [1]{#1}%
\providecommand \bibfnamefont [1]{#1}%
\providecommand \citenamefont [1]{#1}%
\providecommand \href@noop [0]{\@secondoftwo}%
\providecommand \href [0]{\begingroup \@sanitize@url \@href}%
\providecommand \@href[1]{\@@startlink{#1}\@@href}%
\providecommand \@@href[1]{\endgroup#1\@@endlink}%
\providecommand \@sanitize@url [0]{\catcode `\\12\catcode `\$12\catcode
  `\&12\catcode `\#12\catcode `\^12\catcode `\_12\catcode `\%12\relax}%
\providecommand \@@startlink[1]{}%
\providecommand \@@endlink[0]{}%
\providecommand \url  [0]{\begingroup\@sanitize@url \@url }%
\providecommand \@url [1]{\endgroup\@href {#1}{\urlprefix }}%
\providecommand \urlprefix  [0]{URL }%
\providecommand \Eprint [0]{\href }%
\providecommand \doibase [0]{http://dx.doi.org/}%
\providecommand \selectlanguage [0]{\@gobble}%
\providecommand \bibinfo  [0]{\@secondoftwo}%
\providecommand \bibfield  [0]{\@secondoftwo}%
\providecommand \translation [1]{[#1]}%
\providecommand \BibitemOpen [0]{}%
\providecommand \bibitemStop [0]{}%
\providecommand \bibitemNoStop [0]{.\EOS\space}%
\providecommand \EOS [0]{\spacefactor3000\relax}%
\providecommand \BibitemShut  [1]{\csname bibitem#1\endcsname}%
\let\auto@bib@innerbib\@empty
%</preamble>
\bibitem [{\citenamefont {Chowdhury}\ \emph {et~al.}(2005)\citenamefont
  {Chowdhury}, \citenamefont {Schadschneider},\ and\ \citenamefont
  {Nishinari}}]{Chowdhury2005}%
  \BibitemOpen
  \bibfield  {author} {\bibinfo {author} {\bibfnamefont {D.}~\bibnamefont
  {Chowdhury}}, \bibinfo {author} {\bibfnamefont {A.}~\bibnamefont
  {Schadschneider}}, \ and\ \bibinfo {author} {\bibfnamefont {K.}~\bibnamefont
  {Nishinari}},\ }\href
  {http://www.sciencedirect.com/science/article/pii/S1571064505000321}
  {\bibfield  {journal} {\bibinfo  {journal} {Phys. Life Rev.}\ }\textbf
  {\bibinfo {volume} {2}},\ \bibinfo {pages} {318} (\bibinfo {year}
  {2005})}\BibitemShut {NoStop}%
\bibitem [{\citenamefont {Bressloff}\ and\ \citenamefont
  {Newby}(2013)}]{Bressloff2013}%
  \BibitemOpen
  \bibfield  {author} {\bibinfo {author} {\bibfnamefont {P.~C.}\ \bibnamefont
  {Bressloff}}\ and\ \bibinfo {author} {\bibfnamefont {J.~M.}\ \bibnamefont
  {Newby}},\ }\href {http://link.aps.org/doi/10.1103/RevModPhys.85.135}
  {\bibfield  {journal} {\bibinfo  {journal} {Rev. Mod. Phys.}\ }\textbf
  {\bibinfo {volume} {85}},\ \bibinfo {pages} {135} (\bibinfo {year}
  {2013})}\BibitemShut {NoStop}%
\bibitem [{\citenamefont {Neri}\ \emph {et~al.}(2013)\citenamefont {Neri},
  \citenamefont {Kern},\ and\ \citenamefont {Parmeggiani}}]{Neri2013b}%
  \BibitemOpen
  \bibfield  {author} {\bibinfo {author} {\bibfnamefont {I.}~\bibnamefont
  {Neri}}, \bibinfo {author} {\bibfnamefont {N.}~\bibnamefont {Kern}}, \ and\
  \bibinfo {author} {\bibfnamefont {A.}~\bibnamefont {Parmeggiani}},\ }\href
  {http://link.aps.org/doi/10.1103/PhysRevLett.110.098102} {\bibfield
  {journal} {\bibinfo  {journal} {Phys. Rev. Lett.}\ }\textbf {\bibinfo
  {volume} {110}},\ \bibinfo {pages} {098102} (\bibinfo {year}
  {2013})}\BibitemShut {NoStop}%
\bibitem [{\citenamefont {Helbing}(2001)}]{Helbing2001}%
  \BibitemOpen
  \bibfield  {author} {\bibinfo {author} {\bibfnamefont {D.}~\bibnamefont
  {Helbing}},\ }\href {http://link.aps.org/doi/10.1103/RevModPhys.73.1067}
  {\bibfield  {journal} {\bibinfo  {journal} {Rev. Mod. Phys.}\ }\textbf
  {\bibinfo {volume} {73}},\ \bibinfo {pages} {1067} (\bibinfo {year}
  {2001})}\BibitemShut {NoStop}%
\bibitem [{\citenamefont {Chowdhury}\ \emph {et~al.}(2000)\citenamefont
  {Chowdhury}, \citenamefont {Santen},\ and\ \citenamefont
  {Schadschneider}}]{Chowdhury2000}%
  \BibitemOpen
  \bibfield  {author} {\bibinfo {author} {\bibfnamefont {D.}~\bibnamefont
  {Chowdhury}}, \bibinfo {author} {\bibfnamefont {L.}~\bibnamefont {Santen}}, \
  and\ \bibinfo {author} {\bibfnamefont {A.}~\bibnamefont {Schadschneider}},\
  }\href {http://www.sciencedirect.com/science/article/pii/S0370157399001179}
  {\bibfield  {journal} {\bibinfo  {journal} {Phys. Rep.}\ }\textbf {\bibinfo
  {volume} {329}},\ \bibinfo {pages} {199} (\bibinfo {year}
  {2000})}\BibitemShut {NoStop}%
\bibitem [{\citenamefont {Embley}\ \emph {et~al.}(2009)\citenamefont {Embley},
  \citenamefont {Parmeggiani},\ and\ \citenamefont {Kern}}]{Embley2009}%
  \BibitemOpen
  \bibfield  {author} {\bibinfo {author} {\bibfnamefont {B.}~\bibnamefont
  {Embley}}, \bibinfo {author} {\bibfnamefont {A.}~\bibnamefont {Parmeggiani}},
  \ and\ \bibinfo {author} {\bibfnamefont {N.}~\bibnamefont {Kern}},\ }\href
  {http://link.aps.org/doi/10.1103/PhysRevE.80.041128} {\bibfield  {journal}
  {\bibinfo  {journal} {Phys. Rev. E}\ }\textbf {\bibinfo {volume} {80}},\
  \bibinfo {pages} {041128} (\bibinfo {year} {2009})}\BibitemShut {NoStop}%
\bibitem [{\citenamefont {Schadschneider}(2000)}]{Schadschneider2000}%
  \BibitemOpen
  \bibfield  {author} {\bibinfo {author} {\bibfnamefont {A.}~\bibnamefont
  {Schadschneider}},\ }\href {\doibase
  http://dx.doi.org/10.1016/S0378-4371(00)00274-0} {\bibfield  {journal}
  {\bibinfo  {journal} {Physica A}\ }\textbf {\bibinfo {volume} {285}},\
  \bibinfo {pages} {101} (\bibinfo {year} {2000})}\BibitemShut {NoStop}%
\bibitem [{\citenamefont {Forster}\ \emph {et~al.}(1976)\citenamefont
  {Forster}, \citenamefont {Nelson},\ and\ \citenamefont
  {Stephen}}]{Forster1976}%
  \BibitemOpen
  \bibfield  {author} {\bibinfo {author} {\bibfnamefont {D.}~\bibnamefont
  {Forster}}, \bibinfo {author} {\bibfnamefont {D.~R.}\ \bibnamefont {Nelson}},
  \ and\ \bibinfo {author} {\bibfnamefont {M.~J.}\ \bibnamefont {Stephen}},\
  }\href {\doibase 10.1103/PhysRevLett.36.867} {\bibfield  {journal} {\bibinfo
  {journal} {Phys. Rev. Lett.}\ }\textbf {\bibinfo {volume} {36}},\ \bibinfo
  {pages} {867} (\bibinfo {year} {1976})}\BibitemShut {NoStop}%
\bibitem [{\citenamefont {Kardar}\ \emph {et~al.}(1986)\citenamefont {Kardar},
  \citenamefont {Parisi},\ and\ \citenamefont {Zhang}}]{Kardar1986}%
  \BibitemOpen
  \bibfield  {author} {\bibinfo {author} {\bibfnamefont {M.}~\bibnamefont
  {Kardar}}, \bibinfo {author} {\bibfnamefont {G.}~\bibnamefont {Parisi}}, \
  and\ \bibinfo {author} {\bibfnamefont {Y.-C.}\ \bibnamefont {Zhang}},\ }\href
  {http://link.aps.org/doi/10.1103/PhysRevLett.56.889} {\bibfield  {journal}
  {\bibinfo  {journal} {Phys. Rev. Lett.}\ }\textbf {\bibinfo {volume} {56}},\
  \bibinfo {pages} {889} (\bibinfo {year} {1986})}\BibitemShut {NoStop}%
\bibitem [{\citenamefont {Corwin}(2012)}]{Corwin2011}%
  \BibitemOpen
  \bibfield  {author} {\bibinfo {author} {\bibfnamefont {I.}~\bibnamefont
  {Corwin}},\ }\href {\doibase 10.1142/S2010326311300014} {\bibfield  {journal}
  {\bibinfo  {journal} {Rand. Mat.}\ }\textbf {\bibinfo {volume} {01}},\
  \bibinfo {pages} {1130001} (\bibinfo {year} {2012})}\BibitemShut {NoStop}%
\bibitem [{\citenamefont {Quastel}(2012)}]{Quastel2011}%
  \BibitemOpen
  \bibfield  {author} {\bibinfo {author} {\bibfnamefont {J.}~\bibnamefont
  {Quastel}},\ }\href {\doibase 10.4310/CDM.2011.v2011.n1.a3} {\bibfield
  {journal} {\bibinfo  {journal} {Curr. Dev. Math.}\ }\textbf {\bibinfo
  {volume} {2011}},\ \bibinfo {pages} {125} (\bibinfo {year}
  {2012})}\BibitemShut {NoStop}%
\bibitem [{\citenamefont {Derrida}\ \emph {et~al.}(1992)\citenamefont
  {Derrida}, \citenamefont {Domany},\ and\ \citenamefont
  {Mukamel}}]{Derrida1992}%
  \BibitemOpen
  \bibfield  {author} {\bibinfo {author} {\bibfnamefont {B.}~\bibnamefont
  {Derrida}}, \bibinfo {author} {\bibfnamefont {E.}~\bibnamefont {Domany}}, \
  and\ \bibinfo {author} {\bibfnamefont {D.}~\bibnamefont {Mukamel}},\ }\href
  {http://dx.doi.org/10.1007/BF01050430} {\bibfield  {journal} {\bibinfo
  {journal} {J. Stat. Phys.}\ }\textbf {\bibinfo {volume} {69}},\ \bibinfo
  {pages} {667} (\bibinfo {year} {1992})}\BibitemShut {NoStop}%
\bibitem [{\citenamefont {Blythe}\ and\ \citenamefont
  {Evans}(2007)}]{Blythe2007}%
  \BibitemOpen
  \bibfield  {author} {\bibinfo {author} {\bibfnamefont {R.~A.}\ \bibnamefont
  {Blythe}}\ and\ \bibinfo {author} {\bibfnamefont {M.~R.}\ \bibnamefont
  {Evans}},\ }\href {http://stacks.iop.org/1751-8121/40/i=46/a=R01} {\bibfield
  {journal} {\bibinfo  {journal} {J. Phys. A}\ }\textbf {\bibinfo {volume}
  {40}},\ \bibinfo {pages} {R333} (\bibinfo {year} {2007})}\BibitemShut
  {NoStop}%
\bibitem [{\citenamefont {Ha}\ \emph {et~al.}(2003)\citenamefont {Ha},
  \citenamefont {Timonen},\ and\ \citenamefont {den Nijs}}]{MHa2003}%
  \BibitemOpen
  \bibfield  {author} {\bibinfo {author} {\bibfnamefont {M.}~\bibnamefont
  {Ha}}, \bibinfo {author} {\bibfnamefont {J.}~\bibnamefont {Timonen}}, \ and\
  \bibinfo {author} {\bibfnamefont {M.}~\bibnamefont {den Nijs}},\ }\href
  {http://link.aps.org/doi/10.1103/PhysRevE.68.056122} {\bibfield  {journal}
  {\bibinfo  {journal} {Phys. Rev. E}\ }\textbf {\bibinfo {volume} {68}},\
  \bibinfo {pages} {056122} (\bibinfo {year} {2003})}\BibitemShut {NoStop}%
\bibitem [{\citenamefont {Costin}\ \emph {et~al.}(2013)\citenamefont {Costin},
  \citenamefont {Lebowitz}, \citenamefont {Speer},\ and\ \citenamefont
  {Troiani}}]{Costin2012}%
  \BibitemOpen
  \bibfield  {author} {\bibinfo {author} {\bibfnamefont {O.}~\bibnamefont
  {Costin}}, \bibinfo {author} {\bibfnamefont {J.~L.}\ \bibnamefont
  {Lebowitz}}, \bibinfo {author} {\bibfnamefont {E.~R.}\ \bibnamefont {Speer}},
  \ and\ \bibinfo {author} {\bibfnamefont {A.}~\bibnamefont {Troiani}},\ }\href
  {http://web.math.sinica.edu.tw/bulletin/archives_articlecontent16.jsp?bid=MjAxMzEwMw==}
  {\bibfield  {journal} {\bibinfo  {journal} {Bull. Inst. Math., Acad. Sin.
  (New Series)}\ }\textbf {\bibinfo {volume} {8}},\ \bibinfo {pages} {49}
  (\bibinfo {year} {2013})}\BibitemShut {NoStop}%
\bibitem [{\citenamefont {Schmidt}\ \emph {et~al.}(2015)\citenamefont
  {Schmidt}, \citenamefont {Popkov},\ and\ \citenamefont
  {Schadschneider}}]{Schmidt2015}%
  \BibitemOpen
  \bibfield  {author} {\bibinfo {author} {\bibfnamefont {J.}~\bibnamefont
  {Schmidt}}, \bibinfo {author} {\bibfnamefont {V.}~\bibnamefont {Popkov}}, \
  and\ \bibinfo {author} {\bibfnamefont {A.}~\bibnamefont {Schadschneider}},\
  }\href {http://stacks.iop.org/0295-5075/110/i=2/a=20008} {\bibfield
  {journal} {\bibinfo  {journal} {Europhys. Lett.}\ }\textbf {\bibinfo {volume}
  {110}},\ \bibinfo {pages} {20008} (\bibinfo {year} {2015})}\BibitemShut
  {NoStop}%
\bibitem [{\citenamefont {Soh}\ \emph {et~al.}(2017)\citenamefont {Soh},
  \citenamefont {Baek}, \citenamefont {Ha},\ and\ \citenamefont
  {Jeong}}]{HSoh2017}%
  \BibitemOpen
  \bibfield  {author} {\bibinfo {author} {\bibfnamefont {H.}~\bibnamefont
  {Soh}}, \bibinfo {author} {\bibfnamefont {Y.}~\bibnamefont {Baek}}, \bibinfo
  {author} {\bibfnamefont {M.}~\bibnamefont {Ha}}, \ and\ \bibinfo {author}
  {\bibfnamefont {H.}~\bibnamefont {Jeong}},\ }\href {\doibase
  10.1103/PhysRevE.95.042123} {\bibfield  {journal} {\bibinfo  {journal} {Phys.
  Rev. E}\ }\textbf {\bibinfo {volume} {95}},\ \bibinfo {pages} {042123}
  (\bibinfo {year} {2017})}\BibitemShut {NoStop}%
\bibitem [{\citenamefont {Myllys}\ \emph {et~al.}(2003)\citenamefont {Myllys},
  \citenamefont {Maunuksela}, \citenamefont {Merikoski}, \citenamefont
  {Timonen}, \citenamefont {Horv{\'a}th}, \citenamefont {Ha},\ and\
  \citenamefont {den Nijs}}]{Myllys2003}%
  \BibitemOpen
  \bibfield  {author} {\bibinfo {author} {\bibfnamefont {M.}~\bibnamefont
  {Myllys}}, \bibinfo {author} {\bibfnamefont {J.}~\bibnamefont {Maunuksela}},
  \bibinfo {author} {\bibfnamefont {J.}~\bibnamefont {Merikoski}}, \bibinfo
  {author} {\bibfnamefont {J.}~\bibnamefont {Timonen}}, \bibinfo {author}
  {\bibfnamefont {V.~K.}\ \bibnamefont {Horv{\'a}th}}, \bibinfo {author}
  {\bibfnamefont {M.}~\bibnamefont {Ha}}, \ and\ \bibinfo {author}
  {\bibfnamefont {M.}~\bibnamefont {den Nijs}},\ }\href
  {http://link.aps.org/doi/10.1103/PhysRevE.68.051103} {\bibfield  {journal}
  {\bibinfo  {journal} {Phys. Rev. E}\ }\textbf {\bibinfo {volume} {68}},\
  \bibinfo {pages} {051103} (\bibinfo {year} {2003})}\BibitemShut {NoStop}%
\bibitem [{\citenamefont {Kandel}\ and\ \citenamefont
  {Mukamel}(1992)}]{Kandel1992}%
  \BibitemOpen
  \bibfield  {author} {\bibinfo {author} {\bibfnamefont {D.}~\bibnamefont
  {Kandel}}\ and\ \bibinfo {author} {\bibfnamefont {D.}~\bibnamefont
  {Mukamel}},\ }\href {http://dx.doi.org/10.1209/0295-5075/20/4/007} {\bibfield
   {journal} {\bibinfo  {journal} {Europhys. Lett.}\ }\textbf {\bibinfo
  {volume} {20}},\ \bibinfo {pages} {325} (\bibinfo {year} {1992})}\BibitemShut
  {NoStop}%
\bibitem [{\citenamefont {Song}\ and\ \citenamefont {Kim}(2006)}]{HSSong2006}%
  \BibitemOpen
  \bibfield  {author} {\bibinfo {author} {\bibfnamefont {H.~S.}\ \bibnamefont
  {Song}}\ and\ \bibinfo {author} {\bibfnamefont {J.~M.}\ \bibnamefont {Kim}},\
  }\href {http://http://www.jkps.or.kr/journal/view.html?uid=7697&vmd=Full}
  {\bibfield  {journal} {\bibinfo  {journal} {J. Korean. Phy. Soc.}\ }\textbf
  {\bibinfo {volume} {48}},\ \bibinfo {pages} {245} (\bibinfo {year}
  {2006})}\BibitemShut {NoStop}%
\bibitem [{\citenamefont {Lee}\ and\ \citenamefont {Kim}(2009)}]{JHLee2009}%
  \BibitemOpen
  \bibfield  {author} {\bibinfo {author} {\bibfnamefont {J.~H.}\ \bibnamefont
  {Lee}}\ and\ \bibinfo {author} {\bibfnamefont {J.~M.}\ \bibnamefont {Kim}},\
  }\href {http://link.aps.org/doi/10.1103/PhysRevE.79.051127} {\bibfield
  {journal} {\bibinfo  {journal} {Phys. Rev. E}\ }\textbf {\bibinfo {volume}
  {79}},\ \bibinfo {pages} {051127} (\bibinfo {year} {2009})}\BibitemShut
  {NoStop}%
\bibitem [{\citenamefont {{Basu}}\ \emph {et~al.}(2014)\citenamefont {{Basu}},
  \citenamefont {{Sidoravicius}},\ and\ \citenamefont {{Sly}}}]{Basu2014}%
  \BibitemOpen
  \bibfield  {author} {\bibinfo {author} {\bibfnamefont {R.}~\bibnamefont
  {{Basu}}}, \bibinfo {author} {\bibfnamefont {V.}~\bibnamefont
  {{Sidoravicius}}}, \ and\ \bibinfo {author} {\bibfnamefont {A.}~\bibnamefont
  {{Sly}}},\ }\href@noop {} {\  (\bibinfo {year} {2014})},\ \Eprint
  {http://arxiv.org/abs/1408.3464} {arXiv:1408.3464 [math.PR]} \BibitemShut
  {NoStop}%
\bibitem [{\citenamefont {Neri}\ \emph {et~al.}(2011)\citenamefont {Neri},
  \citenamefont {Kern},\ and\ \citenamefont {Parmeggiani}}]{Neri2011}%
  \BibitemOpen
  \bibfield  {author} {\bibinfo {author} {\bibfnamefont {I.}~\bibnamefont
  {Neri}}, \bibinfo {author} {\bibfnamefont {N.}~\bibnamefont {Kern}}, \ and\
  \bibinfo {author} {\bibfnamefont {A.}~\bibnamefont {Parmeggiani}},\ }\href
  {http://link.aps.org/doi/10.1103/PhysRevLett.107.068702} {\bibfield
  {journal} {\bibinfo  {journal} {Phys. Rev. Lett.}\ }\textbf {\bibinfo
  {volume} {107}},\ \bibinfo {pages} {068702} (\bibinfo {year}
  {2011})}\BibitemShut {NoStop}%
\bibitem [{\citenamefont {Baek}\ \emph {et~al.}(2014)\citenamefont {Baek},
  \citenamefont {Ha},\ and\ \citenamefont {Jeong}}]{YBaek2014}%
  \BibitemOpen
  \bibfield  {author} {\bibinfo {author} {\bibfnamefont {Y.}~\bibnamefont
  {Baek}}, \bibinfo {author} {\bibfnamefont {M.}~\bibnamefont {Ha}}, \ and\
  \bibinfo {author} {\bibfnamefont {H.}~\bibnamefont {Jeong}},\ }\href
  {\doibase 10.1103/PhysRevE.90.062111} {\bibfield  {journal} {\bibinfo
  {journal} {Phys. Rev. E}\ }\textbf {\bibinfo {volume} {90}},\ \bibinfo
  {pages} {062111} (\bibinfo {year} {2014})}\BibitemShut {NoStop}%
\bibitem [{\citenamefont {Spitzer}(1970)}]{Spitzer1970}%
  \BibitemOpen
  \bibfield  {author} {\bibinfo {author} {\bibfnamefont {F.}~\bibnamefont
  {Spitzer}},\ }\href {\doibase https://doi.org/10.1016/0001-8708(70)90034-4}
  {\bibfield  {journal} {\bibinfo  {journal} {Adv. Math.}\ }\textbf {\bibinfo
  {volume} {5}},\ \bibinfo {pages} {246 } (\bibinfo {year} {1970})}\BibitemShut
  {NoStop}%
\bibitem [{\citenamefont {Godreche}(2003)}]{Godreche2003}%
  \BibitemOpen
  \bibfield  {author} {\bibinfo {author} {\bibfnamefont {C.}~\bibnamefont
  {Godreche}},\ }\href {http://stacks.iop.org/0305-4470/36/i=23/a=303}
  {\bibfield  {journal} {\bibinfo  {journal} {J. Phys. A: Math. Gen.}\ }\textbf
  {\bibinfo {volume} {36}},\ \bibinfo {pages} {6313} (\bibinfo {year}
  {2003})}\BibitemShut {NoStop}%
\bibitem [{\citenamefont {Evans}\ and\ \citenamefont
  {Hanney}(2003)}]{Evans2003}%
  \BibitemOpen
  \bibfield  {author} {\bibinfo {author} {\bibfnamefont {M.~R.}\ \bibnamefont
  {Evans}}\ and\ \bibinfo {author} {\bibfnamefont {T.}~\bibnamefont {Hanney}},\
  }\href {http://stacks.iop.org/0305-4470/36/i=28/a=101} {\bibfield  {journal}
  {\bibinfo  {journal} {J. Phys. A: Math. Gen.}\ }\textbf {\bibinfo {volume}
  {36}},\ \bibinfo {pages} {L441} (\bibinfo {year} {2003})}\BibitemShut
  {NoStop}%
\bibitem [{Note1()}]{Note1}%
  \BibitemOpen
  \bibinfo {note} {The ZRP is an exact mapping of the periodic TASEP without
  the slow bond.}\BibitemShut {Stop}%
\bibitem [{\citenamefont {Janowsky}\ and\ \citenamefont
  {Lebowitz}(1992)}]{Janowsky1992}%
  \BibitemOpen
  \bibfield  {author} {\bibinfo {author} {\bibfnamefont {S.~A.}\ \bibnamefont
  {Janowsky}}\ and\ \bibinfo {author} {\bibfnamefont {J.~L.}\ \bibnamefont
  {Lebowitz}},\ }\href {http://link.aps.org/doi/10.1103/PhysRevA.45.618}
  {\bibfield  {journal} {\bibinfo  {journal} {Phys. Rev. A}\ }\textbf {\bibinfo
  {volume} {45}},\ \bibinfo {pages} {618} (\bibinfo {year} {1992})}\BibitemShut
  {NoStop}%
\bibitem [{\citenamefont {Gro{\ss}kinsky}\ \emph {et~al.}(2003)\citenamefont
  {Gro{\ss}kinsky}, \citenamefont {Sch{\"u}tz},\ and\ \citenamefont
  {Spohn}}]{Grosskinsky2003}%
  \BibitemOpen
  \bibfield  {author} {\bibinfo {author} {\bibfnamefont {S.}~\bibnamefont
  {Gro{\ss}kinsky}}, \bibinfo {author} {\bibfnamefont {G.~M.}\ \bibnamefont
  {Sch{\"u}tz}}, \ and\ \bibinfo {author} {\bibfnamefont {H.}~\bibnamefont
  {Spohn}},\ }\href {\doibase 10.1023/A:1026008532442} {\bibfield  {journal}
  {\bibinfo  {journal} {J. Stat. Phys.}\ }\textbf {\bibinfo {volume} {113}},\
  \bibinfo {pages} {389} (\bibinfo {year} {2003})}\BibitemShut {NoStop}%
\bibitem [{Note2()}]{Note2}%
  \BibitemOpen
  \bibinfo {note} {Previous studies~\cite {MHa2003,HSoh2017} used the average
  particle occupation at $x=L/4$ as a good proxy of the bulk density, since the
  particle-hole symmetry in the original TASEP manifests the boundary of HD/LD
  locates at $x=L/2$. With the broken symmetry ($b \not =0$), the boundary does
  not locate at $x=L/4$.}\BibitemShut {Stop}%
\bibitem [{\citenamefont {Evans}(2000)}]{Evans2000}%
  \BibitemOpen
  \bibfield  {author} {\bibinfo {author} {\bibfnamefont {M.~R.}\ \bibnamefont
  {Evans}},\ }\href
  {http://www.scielo.br/scielo.php?script=sci_arttext&pid=S0103-97332000000100005&nrm=iso}
  {\bibfield  {journal} {\bibinfo  {journal} {Braz. J. Phys.}\ }\textbf
  {\bibinfo {volume} {30}},\ \bibinfo {pages} {42 } (\bibinfo {year}
  {2000})}\BibitemShut {NoStop}%
\bibitem [{Note3()}]{Note3}%
  \BibitemOpen
  \bibinfo {note} {M. Abramowitz. Handbook of Mathematical Functions. Dover,
  New York, 1972.}\BibitemShut {Stop}%
\end{thebibliography}%

\end{document}